\documentclass[manuscript]{aastex}

\usepackage{natbib}

\newcommand{\be}{\begin{equation}}
\newcommand{\ee}{\end{equation}}

\shorttitle{Sign singularity, magnetic field complexity and flares in solar active region NOAA 11158}
\shortauthors{Sorriso-Valvo et al.}

\begin{document}

\title{Sign singularity and flares in solar active region NOAA 11158}

\author{
L. Sorriso-Valvo\altaffilmark{1,2},
G. De Vita\altaffilmark{3},
M. Kazachenko\altaffilmark{2},
S. Krucker\altaffilmark{2,4}
L. Primavera\altaffilmark{3},
S. Servidio\altaffilmark{3},
A. Vecchio\altaffilmark{5},
B. Welsch\altaffilmark{2},
G. Fisher\altaffilmark{2},
F. Lepreti\altaffilmark{3},
V. Carbone\altaffilmark{3}
}

\altaffiltext{1}{IPCF-CNR, U.O. di Cosenza, Ponte P. Bucci, cubo 31C, 87036 Rende, Italy.}
\altaffiltext{2}{Space Sciences Laboratory, University of California, 7 Gauss way, Berkeley 94720, California, USA.}
\altaffiltext{3}{Dipartimento di Fisica, Universit\`a della Calabria, Ponte P. Bucci, cubo 31C, 87036 Rende, Italy.}
\altaffiltext{4}{Institute of 4D Technologies, School of Engineering, University of Applied Sciences and Arts Northwestern Switzerland, 5210 Windisch, Switzerland.}
\altaffiltext{5}{INGV, sede di Cosenza, ponte P. Bucci, cubo 30C, 87036 Rende, Italy.}

\email{sorriso@fis.unical.it}

\begin{abstract}
Solar Active Region NOAA 11158 has hosted a number of strong flares, including one X2.2 event. The current density and current helicity complexity properties are studied through cancellation analysis of their sign-singular measure, which features power-law scaling. Spectral analysis is also performed, revelaing the presence of two separate scaling ranges with diffrent spectral index. The time evolution of parameters is discussed. Sudden changes of the cancellation exponents at the time of large flares, and the presence of correlation with EUV and X-ray flux, suggest that eruption of large flares can be linked to the small scale properties of the current structures. 
\end{abstract}


%
%
%
\section{Introduction}
\label{sec-intro}
Solar magnetic activity is often accompanied by spectacular, abrupt phenomena, such as solar flares~\citep{benz} and coronal mass ejections~\citep{chen}. These eruptive, highly energetic features can produce variations in the Sun-Earth connection, resulting for example in geomagnetic storms and other disturbances that can affect human activities and health~\citep{schwenn,pulkkinen}. The increasing amount and the improved quality of solar observations, both from Earth and from space, has provided enormous advances in the understanding of the physical processes occurring in the solar regions associated with flares, i.e. the solar active regions (ARs). These are regions where emerging photospheric magnetic field concentrates in bi- or multi-polar structures, which may include the presence of pores and sunspots. Driven by convective motions of the external layers of the Sun, active regions entangled and twisted magnetic fields can store a considerable amount of non-potential magnetic energy. The rapid release of such energy, probably due to dissipation in magnetic reconnection, is thought to be the basic mechanism of solar flares~\citep{carmichael,sturrock,hirayama,kopp}. 

One important topic in solar physics is the identification of magnetic signature of the occurrence of flares within ARs. In recent years, convincing evidences of the turbulent nature of the ARs magnetic fields dynamics have emerged~\citep{abra-turbo}. In such context, the complexity of the photospheric magnetic field structure arises from the strongly nonlinear, coupled interactions between the plasma flow and magnetic fluctuations on different scales, and results in the superposition of correlated structures~\citep{frisch,biskampbook}. Typical power-law spectra of photospheric magnetic energy have been reported~\citep{zhang2014}, with spectral indices compatible with a Kolmogorov-type phenomenology~\citep{k41,frisch}. Turbulent photospheric magnetic fields have also been described as intermittent and multifractal~\citep{abra-turbo,abra-inter}, which are typical features of turbulent plasmas. This implies the presence of a hierarchy of correlated fluctuations, which concentrates energy on localized, small scale structures, where enhanced dissipation occurs. 

In order to capture the dynamical properties of the intermittent structures, cancellation analysis has been recently used with the aim of correlating the complexity of solar magnetic fields in active regions to the occurrence of flares~\citep{yurchyshyn,abramenko,prelude,hinode}. Such analysis has highlighted the importance of sign singularities in the energy storage process that could lead to flare eruption. In this paper we use for the first time cancellation analysis to describe the fine time and space resolution dynamics of AR NOAA 11158. Corroborated by the study of other observables, this analysis confirms the presence of nontrivial correlations between the topological changes of magnetic structures and flaring activity.

%

\section{Signed measure and cancellation analysis}
\label{sec-cancellations}

Solar active regions are often characterized by scale dependent formation of energetic and localized magnetic structures~\citep{abramenko}.
Because of their coherence, structures can be seen as smooth regions of the magnetic field, embedded in a highly fluctuating background. For zero-mean fields, they can be associated to scale dependent, signed fluctuations of the fields. By introducing a signed measure (as opposed to the usual positive defined probability measure), it is possible to characterize the scaling properties of sign oscillations (or sign persistence) of the fields~\citep{Ott}. Therefore, the presence and the topological characteristics of structures defined in sign can be studied. Signed measure has been successfully used to describe the cancellation properties of magnetic dynamo~\citep{Ott}, as well as the characteristics of current structures in turbulent magnetohydrodynamic (MHD)~\citep{Luca,Annick}, Hall-MHD~\citep{luis}, and kinetic~\citep{devita,karimabadi} numerical simulations. Applications to measurements of magnetic vectors in solar active regions have confirmed cancellation analysis as an interesting tool to detect changes in the scaling properties of the fields fluctuations, and of the fractal dimension of the associated gradients~\citep{yurchyshyn,abramenko,prelude}. 

The signed measure of a zero-mean scalar field~$f(\bf{r})$ can be defined on a $d$-dimensional set $Q(L)$ of size~$L$ as follows~\citep{Ott}. 
Let $\{Q_i(l) \subset Q(L)\}$ be a partition of $Q(L)$ in disjoint subsets of size~$l$. Then, for each scale~$l$ and for each disjoint set of boxes~$Q_i(l)$, the signed measure is 
  \begin{equation} 
  \mu_i(l)=\frac{\int_{Q_i(l)}\;\mathrm{d}{\bf r}\;f({\bf r})} 
  {\int_{Q(L)}\;\mathrm{d}{\bf r}\;|f({\bf r})|} \ . 
  \label{eq_mu} 
  \end{equation} 
When the size of the subset~$Q_i(l)$ is large, cancellations between small structures of opposite sign occur within each box, resulting in small contribution to the signed measure. However, as the boxes become smaller and reach the typical size of the structures, each box is more likely to contain one single, sign defined structure, reducing the level of cancellations. The way this happens can be statistically characterized through the cancellation function 
  \begin{equation} 
  \chi(l) =\sum_{Q_i(l)} |\mu_i(l)|  
  \label{chi1} 
  \end{equation} 
where the sum is extended to all disjoint subsets~$Q_i(l)$. Contrary to positive defined probability measures, the signed measure holds information on the sign of the field fluctuations. In particular, if the measure changes sign on arbitrarily fine scale  (i.e. if for any subset $Q_A(l)$ for which $\mu_A(l) \ne 0$ there exists a subset  $Q_B(l^\prime) \subset Q_A(l)$ such that $\mu_B(l^\prime)$ has opposite sign from $\mu_A(l)$), then the measure is called {\it sign-singular}~\citep{Ott}. Upon performing a scale dependent partition of the whole domain, the sign-singularity of the measure can be quantitatively estimated through the 
cancellation exponent $\kappa$, that is the scaling exponent of the cancellation function, defined as
  \begin{equation} 
  \chi(l) =\sum_{Q_i(l)} |\mu_i(l)| \sim l^{-\kappa} \ ,
  \label{chi} 
  \end{equation} 
where the sum is extended to all disjoint subsets~$Q_i(l)$ that cover the domain~$Q(L)$. In a fluctuating field with positive and negative structures, cancellations occur in large size subsets, providing small contribution to the signed measure. Conversely, when the subsets become smaller and reach the typical size of the structures, the enhanced presence of sign-defined structures reduces the level of cancellations, and consequently increase the relative contribution to the signed measure. Thus, the cancellation exponent represents an effective measure of the efficiency of the field cancellations. Specific examples are represented by a smooth field (with no sign-singularity), for which the cancellation function has a constant value (so that $\kappa=0$), and by a homogeneous field with random discontinuities (i.e. Brownian noise), for which $\kappa=d/2$. Cancellation exponents between those two limiting values indicate presence of smooth structures embedded in random fluctuations. Moreover, their values can be related to the geometrical properties of structures. Furthermore, it is possible to establish a phenomenological relationship between the cancellation exponent and the fractal dimension $D$ of the typical dissipative structures of the flow~\citep{Luca},    
  \begin{equation}
  \kappa=(d-D)/2 \ .
  \label{fractal}
  \end{equation} 
On the other hand, cancellation exponents larger than the typical value for random fields $\kappa>d/2$ indicate the presence of sign anti-persistent structures. Such behaviour is typically associated to the presence of pairs of adjacent, opposite sign structures, which enhance cancellations with respect to random fluctuations.
It should be pointed out that the use of one single fractal dimension cannot fully capture all the fine details of the typical plasma tubulence processes, which are more likely characterized by multifractal scaling~\citep{biskamp}. Nonetheless, $D$ still represents a useful indicator of the topological characteristics of the ``mean'' intermittent structures of the flow.

%
\section{Solar Active Region NOAA 11158}
\label{sec-data}

\subsection{Data reduction: AR 11158} \label{data} 

We derive evolution of magnetic fields in NOAA 11158 using series of HMI vector magnetograms. A 6-day uninterrupted, 12-minute cadence data-set allowed us to study in detail both the long-term, gradual evolution, as well as the rapid changes during an X-class flare.  In this section we describe the data-set and the re-projection we use.  

NOAA 11158 was the source of an X2.2 flare on 2011/02/15 starting at 01:44 UT, peaking at 01:56 UT and ending at 02:06 UT. A front-side halo CME accompanied the flare~\citep{Schrijver2011}.  Prior to the X2.2 flare, the largest flare in this region was an M6.6 on 2011/02/13 at 17:28 UT, a little more than 30 hours before the flare of study.

Helioseismic and Magnetic Imager (HMI) of NASA's Solar Dynamics Observatory (SDO) satellite~\citep{Pesnell2012} observed the NOAA 11158 in high detail, routinely generating filtergrams in six polarization states at six wavelengths on the Fe~I~617.3~nm spectral line. From these filtergrams, images for the Stokes parameters, I, Q, U, and V were derived which, using the Very Fast Inversion of the Stokes Algorithm (VFISV) code~\citep{Borrero2011}, were inverted into the magnetic field vector components.  To resolve the $180^\circ$ azimuthal field ambiguity we used the ``minimum energy'' method~\citep{Metcalf1994,Leka2009}. In addition, we corrected a few episodes of single-frame fluctuations in the direction of the transverse magnetic field vector, by nearly $180^{\circ}$~\citep{Welsch2013}.

To study pre-flare photospheric magnetic evolution, and to baseline this evolution against post-flare evolution, we retained 153 hours of 12-minute-cadence 0.5''-pixel resolution HMI vector-magnetogram data, from beginning of the active region emergence, around four days before the X2.2 flare, to two days after the flare:  $t_{start}=$ February 10 2011 14:11 UT $(S19, E50)$, $t_{end}=$ February 16 2011 23:35 UT $(S21, W37)$.

To account for the non-zero viewing angle of the observed vector magnetic fields and Doppler velocities, we reprojected the HMI-data cube to the disk center and transformed it to Cartesian coordinates~\citep{Welsch2013}. To do that, in the first step, we re-projected the observed magnetic vectors' components onto radial/horizontal coordinate axes. We then determined the shift of the grid's center between frames. After we determined a shift, we converted and then translated by the decided-upon shift the Cartesian grid's points to LOS/POS coordinates. We then interpolated the radial and horizontal components of the magnetic field, $B_r$, $B_h$, onto the grid. Finally, we interpolated data in LOS/POS coords onto a fixed, Cartesian grid that accounts for shifts. This interpolation implied a re-projection of the magnetogram surface, from spherical to a regular, Cartesian grid. Since we used Fourier Local Correlation Tracking (FLCT) to find velocities, we needed conformal transformation into cartesian coordinates; we chose Mercator re-projection~\citep{Welsch2009}, with regular longitudes/latitudes. After re-projection, to preserve physical quantities of magnetic fields and velocities, we corrected the fluxes for the distortion of pixel areas introduced by re-projection; the details of the applied correction-factors are given in Appendix A of~\citet{Kazachenko2014b}. For the minimum magnetic field to consider, we chose a threshold of $|{\bf B}|=250$~Gauss, consistent with the upper limit of the uncertainty in the horizontal magnetic field~\citep{Liu2012a}. To avoid spurious signals in magnetic fields, we apply a mask, where we set any pixel's magnetic field components to zero, if in any of the three consecutive frames it has $|{\bf B}|<250$~Gauss. We also added a boundary area of 55-pixels width/height of padded with zeroes~\citep{Kazachenko2014a}. The final data cube  after re-projection and boundary adding consists of 768 time steps ($dt=720s$) and has a field of view of $665\times645$ pixels with a pixel size of $360.16$ km, which is equivalent to the original  0.5'' pixels HMI-data resolution. More details on the data cube preparation and calibration, for a shorter time range, could be found in~\cite{Welsch2013}.

\begin{figure*}[ht!]
  \includegraphics[width=15cm]{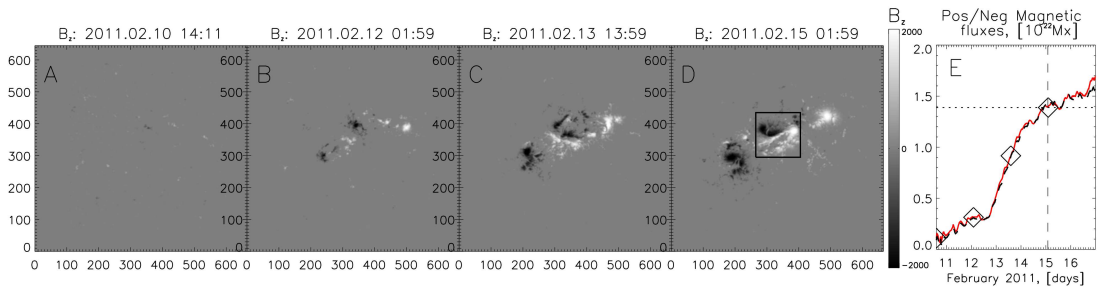}
  \caption{{\it Panels A-D:} HMI vertical magnetic field ($B_z$) maps at 4 different times of AR 11158 evolution. {\it Panel E}: evolution of the positive and negative vertical magnetic fluxes of the 6-day interval, with the diamonds indicating the times of the four images (A-D) on the left. An X-class flare occurred at the time corresponding to the vertical dashed line. The black box in {\it Panel D} indicates the Region where the X2.2 flare was located.}
  \label{fig_mag}
\end{figure*}

Fig.~\ref{fig_mag} shows the final LOS magnetic field in a subregion of the full-disk data array after re-projection in the beginning (Panel $A$), middle ($B,C$) and the post-flare ($D$) times of the magnetogram sequence. Note that the positive and negative magnetic fluxes, shown on the right plot, balance each other; the magnetic flux increases from essentially zero to roughly $1.4\times10^{22}Mx$ at the time of the flare (vertical dashed line).
To summarize, as a result of the data reduction we obtained a data cube consisting of 768 frames, each of which contains data for  three components of the magnetic field with a field of view of $665\times645$ pixels, a pixel size of $360.16$ km and a time step of $dt=720$ sec.

\begin{figure*}[h!] 
  \includegraphics[width=8.2cm]{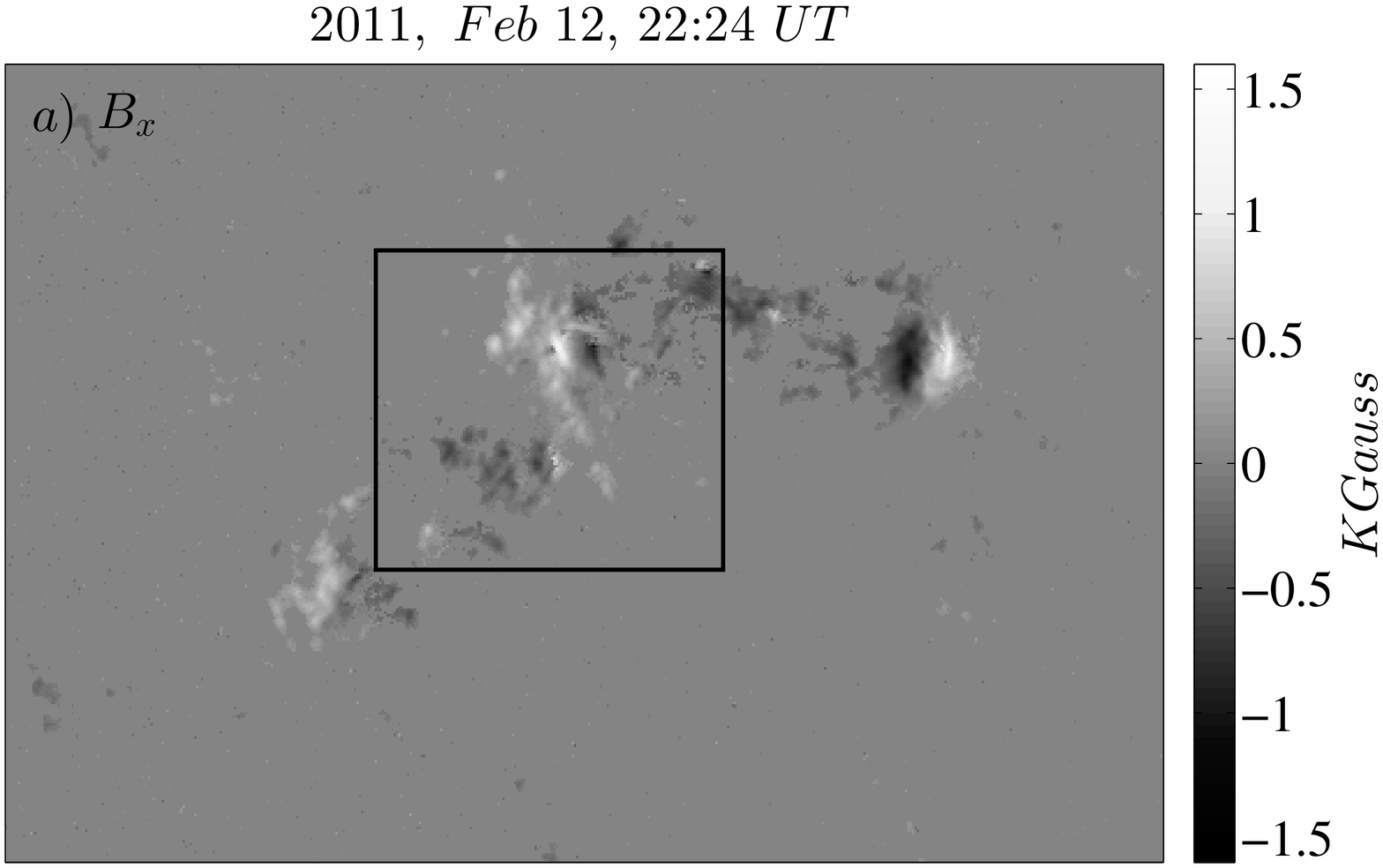}\hskip 10pt \includegraphics[width=8.2cm]{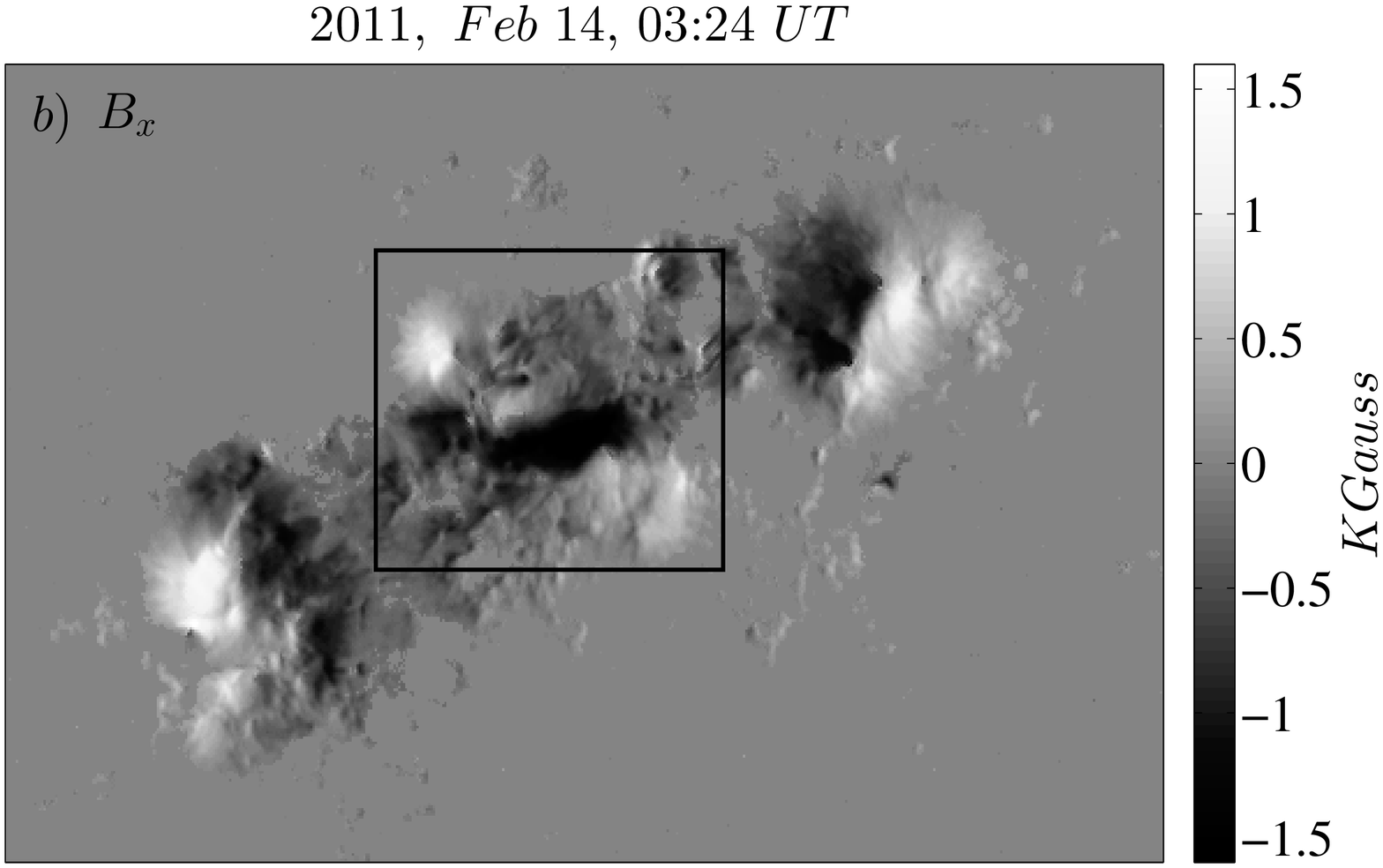}\\
  \includegraphics[width=8.2cm]{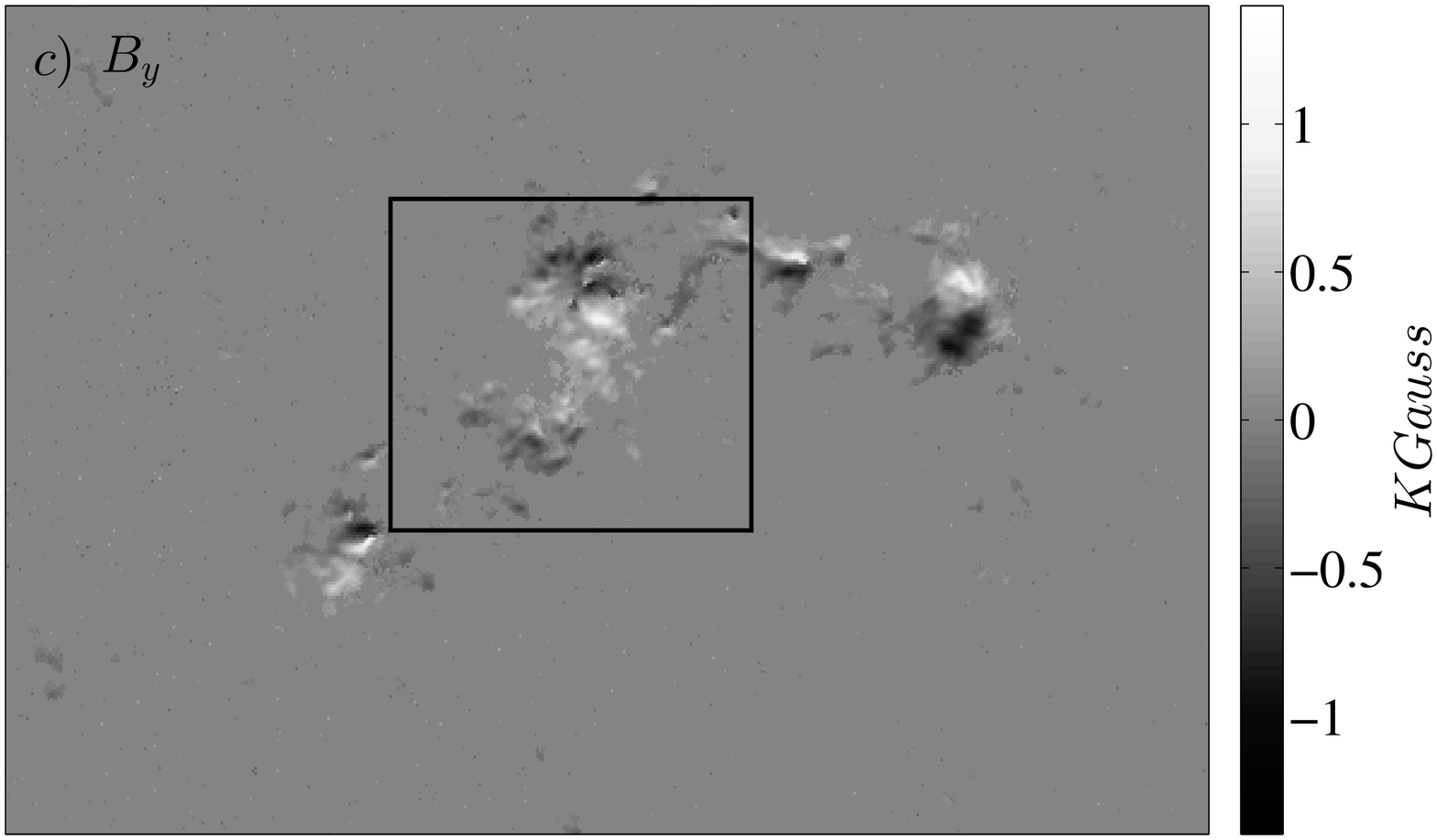}\hskip 10pt \includegraphics[width=8.2cm]{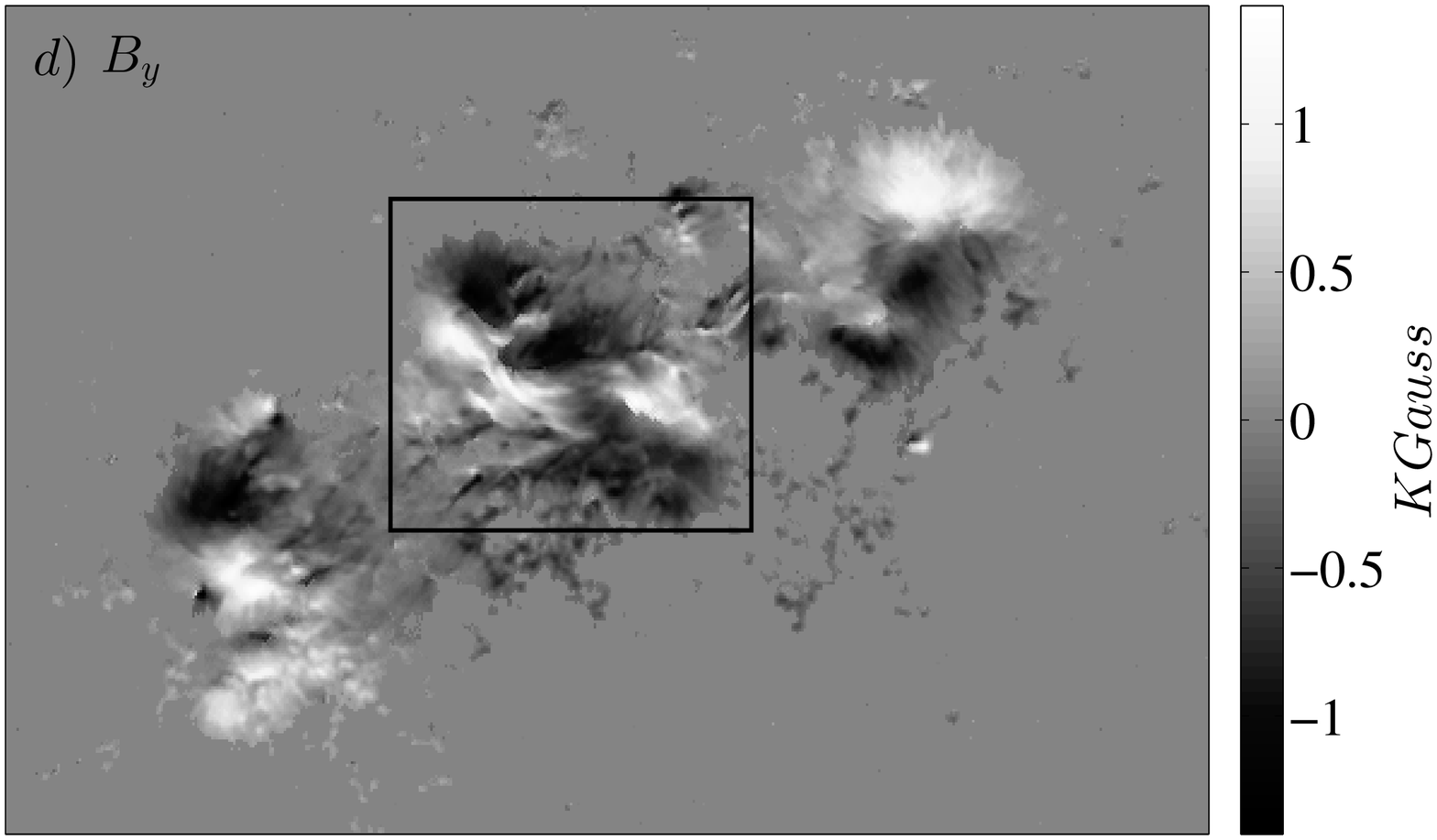}\\
  \includegraphics[width=8cm]{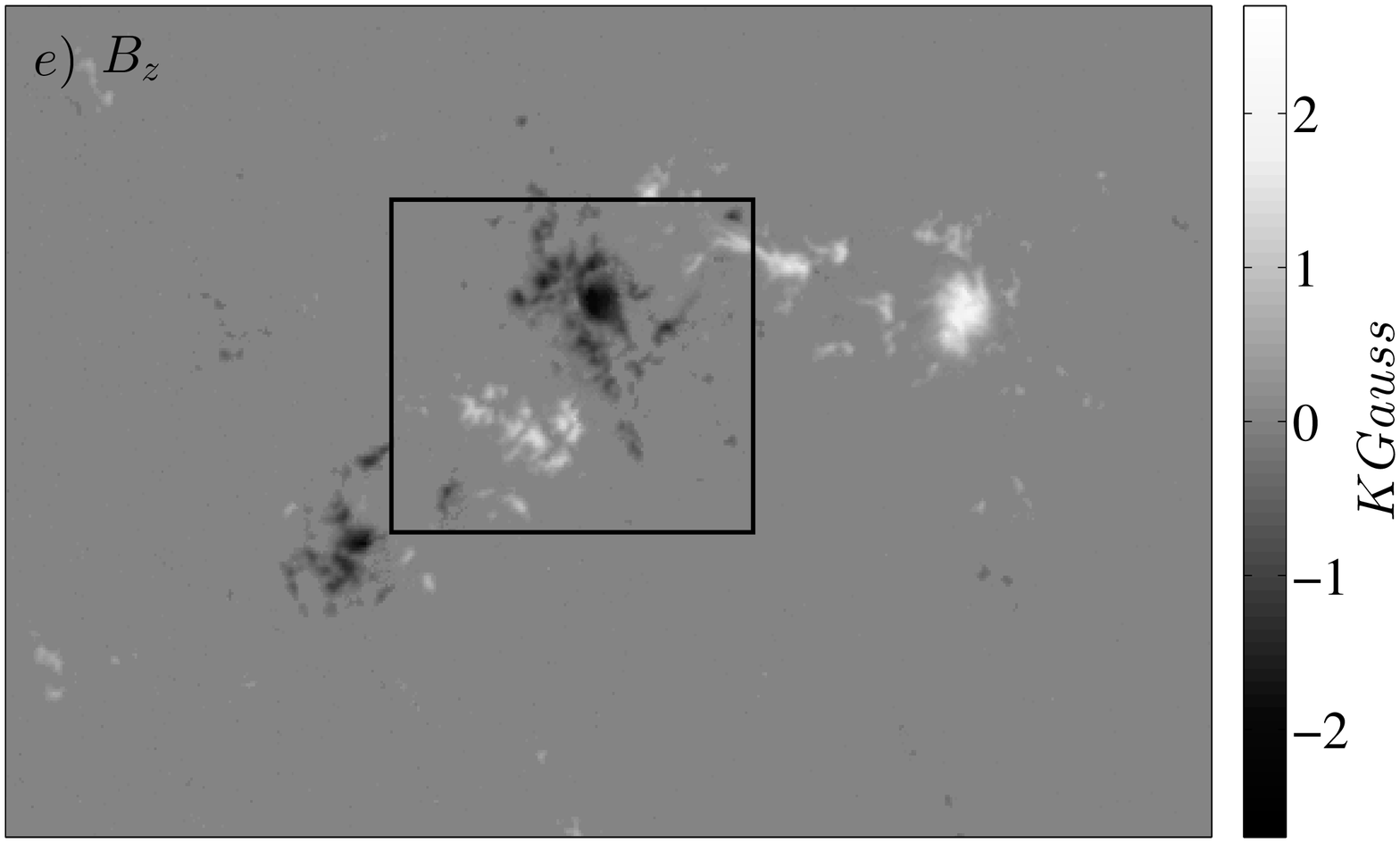}  \hskip 15pt \includegraphics[width=8cm]{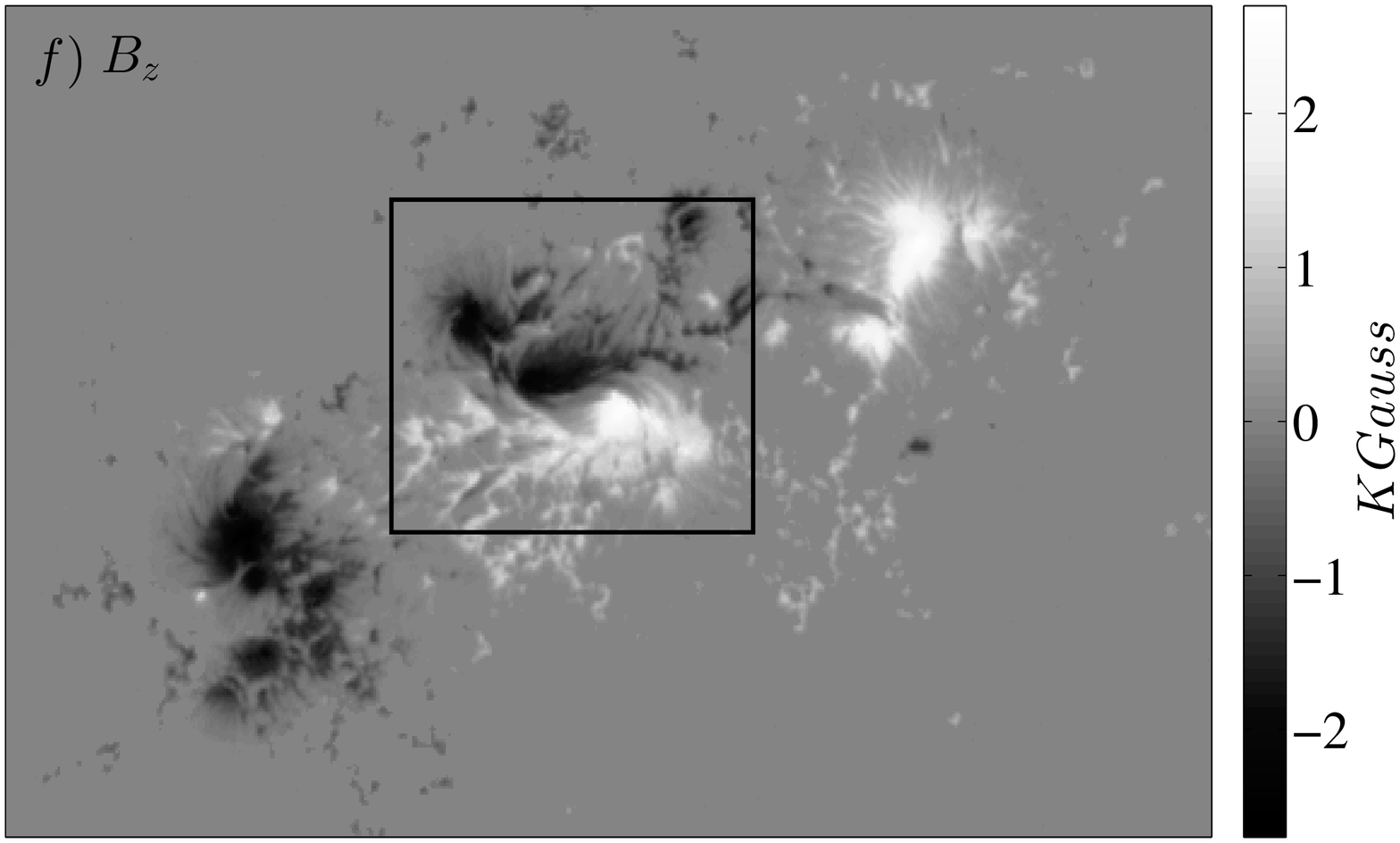}\\
\caption{The magnetic field components intensity measured for NOAA 11158 on February 12th at 22:24UT (left panels) and on February 14th at 03:24UT (right panels).}
  \label{fig_AR}
  \end{figure*}

For sake of complenteness, Figure~\ref{fig_AR} shows the three magnetic field components at two different times, on February 12th at 22:24UT (left panels) and on February 14th at 03:24UT (right panels), before and after the flux emergence. The black solid line box indicates the reduced area used for the cancellation analysis.
From the figure, the evolution of the magnetic structure of the AR is evident. The full time evolution of the magnetic field component $B_z$ is shown in the movie~\ref{movie}, where it is possible to observe the clear flux emergence, and the successive increase of the magnetic complexity.
%
%
%
  \begin{figure*}[h!] 
  \label{movie}
  \caption{A clip of the time evolution of the dataset and relative quantities evaluated in this paper.}
  \end{figure*}
%

\subsection{Preliminary analysis and complementary measurements}

\begin{figure*}[h!] 
  \includegraphics[width=8cm]{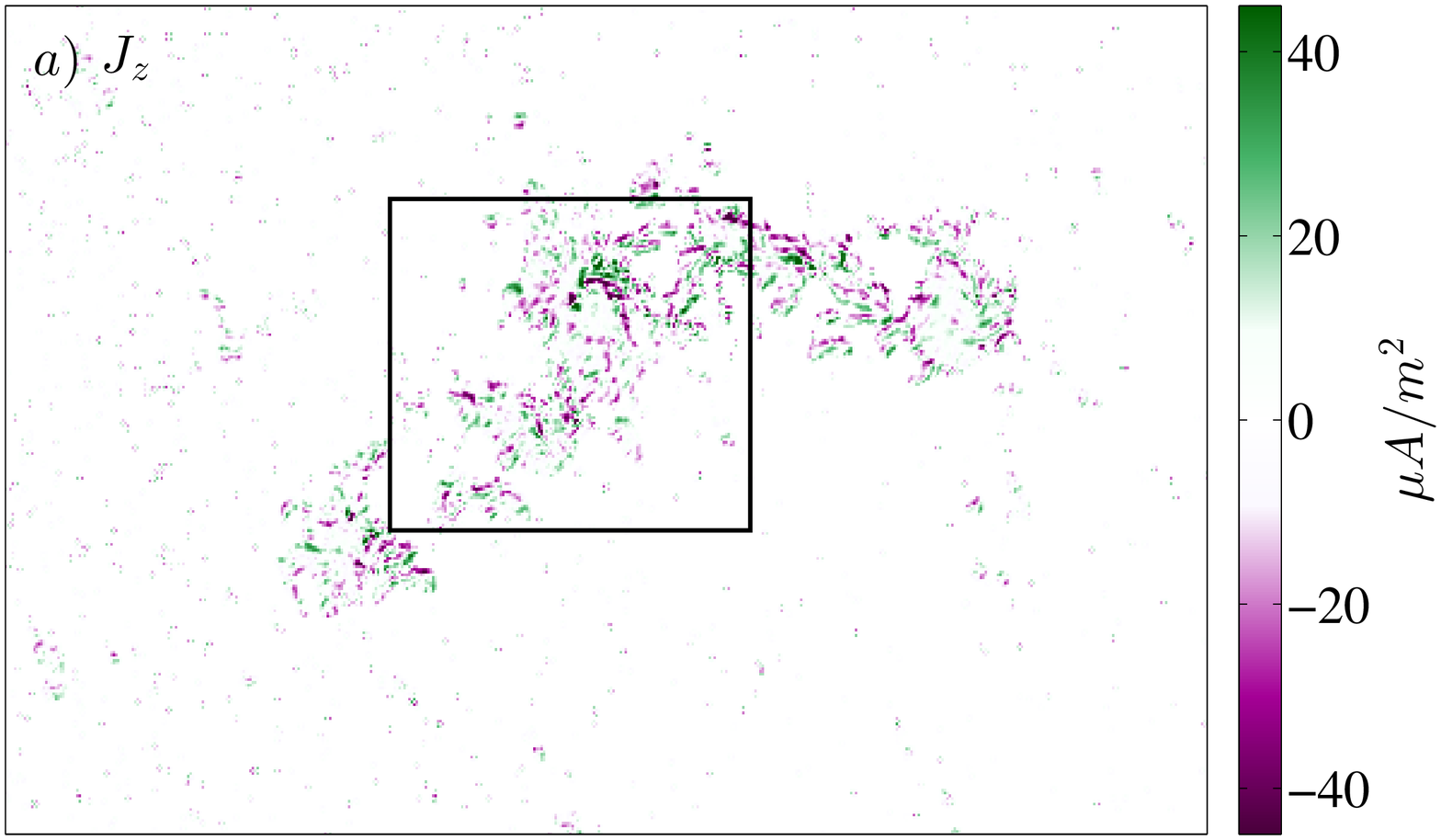} \hskip 15pt\includegraphics[width=8cm]{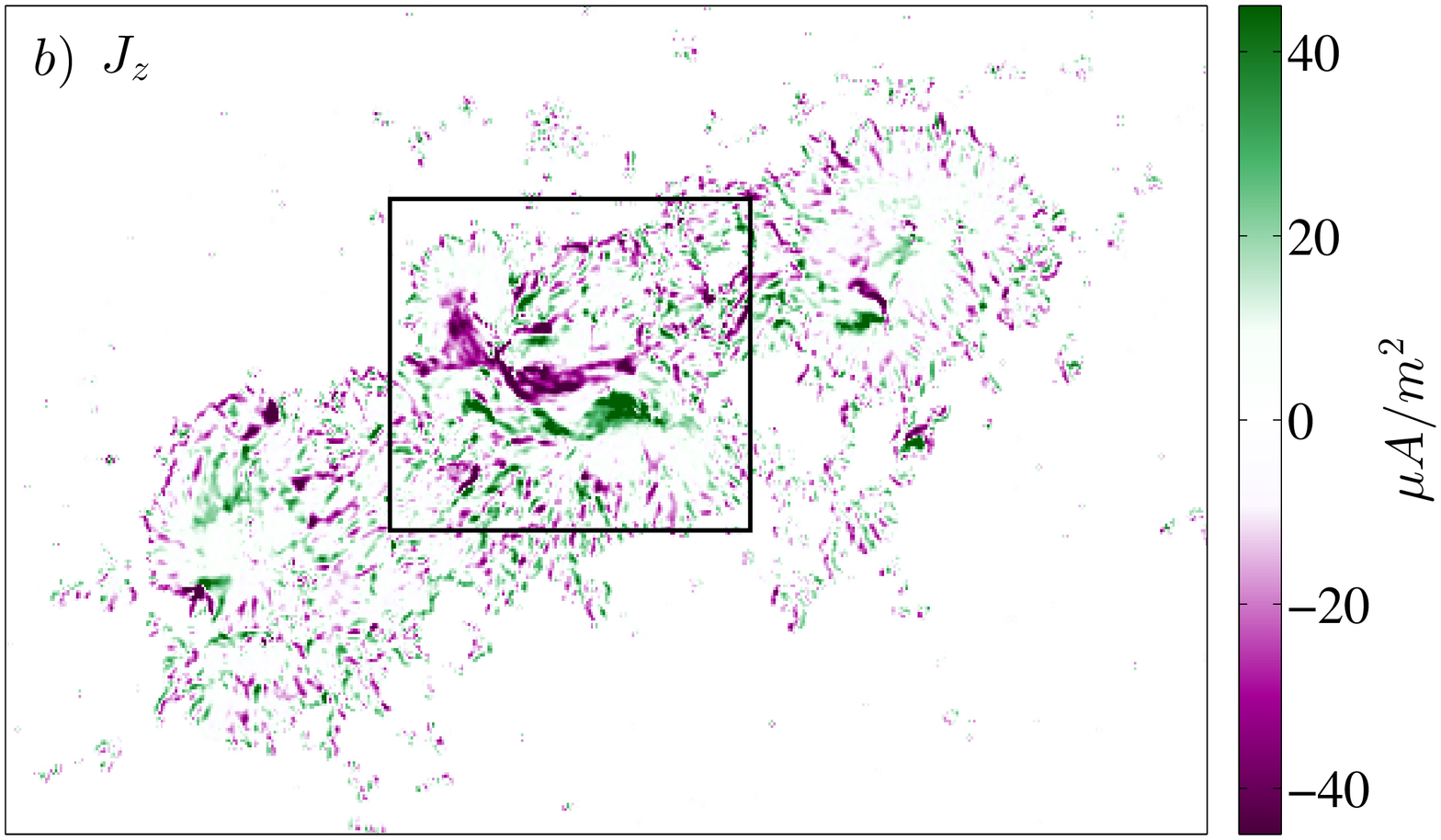}\\
  \includegraphics[width=8.2cm]{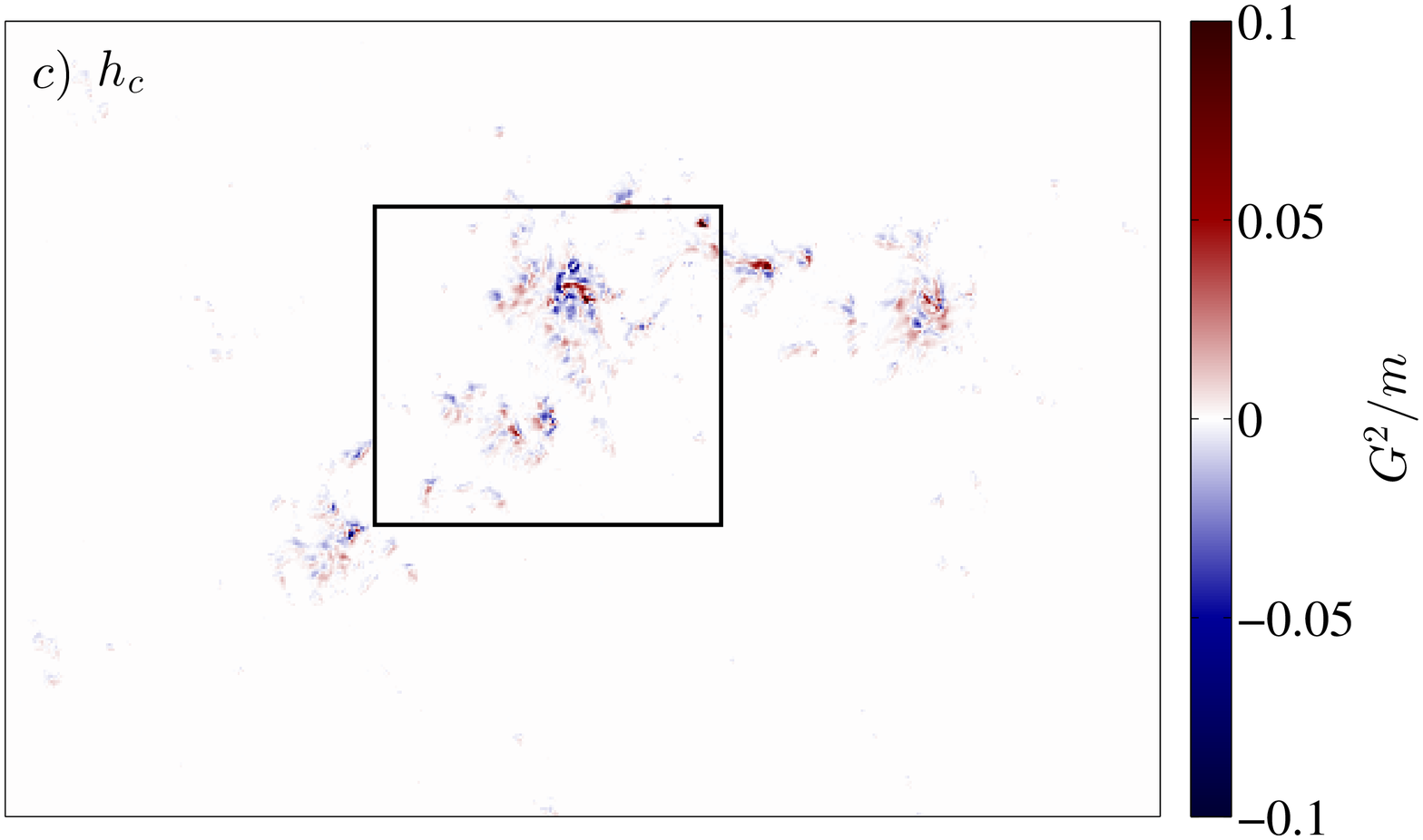} \hskip 10pt \includegraphics[width=8.2cm]{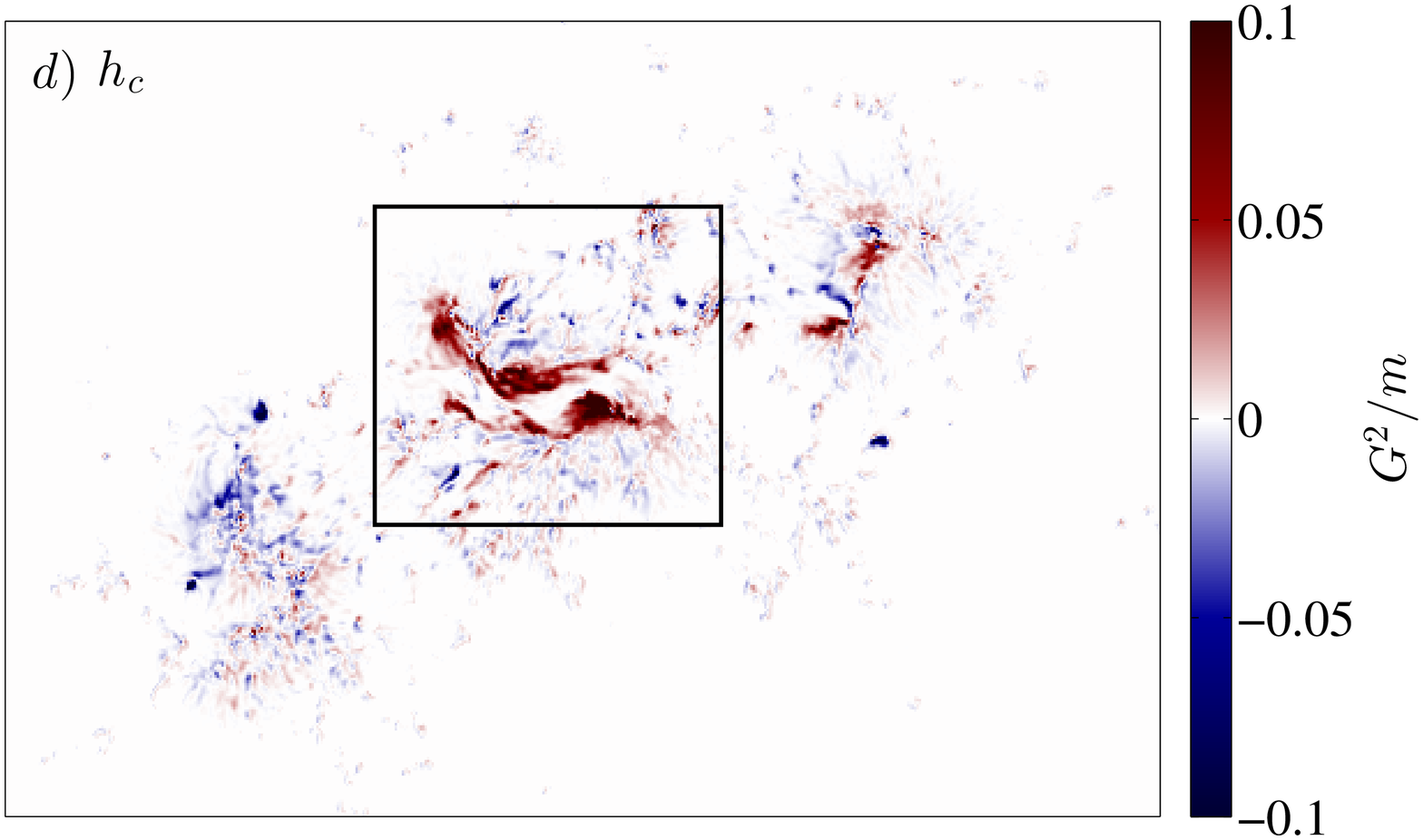}\\    
\caption{The vertical component of the current density $J_z$ (panels a, b) and the reduced current helicity $h_c$ (panels c, d), computed for NOAA 11158 on February 12th at 22:24UT (left panels) and on February 14th at 03:24UT (right panels).}
  \label{fig_j_hc}
  \end{figure*}

The measurements of the magnetic field vector allow to calculate the current density component perpendicular to the solar surface, $J_z$, shown in panels (a) and (b) of Figure~\ref{fig_j_hc} at the two times presented in Figure~\ref{fig_AR}. 
The vertical component of the current density $J_z(x,y)$ has been estimated through the photospheric vector magnetic field $\mathbf{B}(x,y)$, where $(x,y)$ are the cartesian coordinates on the solar surface, as the line integral of the transversal component of the magnetic field over a closed contour $G$, $4\pi J_z/c = (\mathbf{\nabla\times B})_z = s^{-1} \oint_G \mathbf{B}_\perp \cdot d \mathbf{r}$. The integral along each pixel $G$ of area $s=(0.361\times0.361)Mm^2$ was computed using Simpson's rule. 
An alternative calculation, based on the finite differences evaluation of the magnetic field rotational, has been also performed, providing identical results. 
Previous studies of cancellations in solar AR have shown that it is convenient to use the reduced current helicity $h_c=J_zB_z$~\citep{abramenko,jing2012}. This field carries important information about the non-potential magnetic energy available in the AR. Furthermore, it is normally less noisy than the current density, so that cancellation effects are easier to identify. As a matter of fact, previous analyses were unable to identify sign-singularities in the current density, which were only observed in current helicity. It should be pointed out that, since it is not possible to measure the full current density vector with the available magnetic field measurements, in this work we limit our discussion to the current vertical component and the reduced current helicity  $h_c(x,y)=B_z(x,y)J_z(x,y)$, and assume that they are approximating the corresponding full quantities. 
Panels (c) and (d) of figure~\ref{fig_j_hc} show, for the same two snapshots as in previous figures, the estimated current helicity, which has a smoother appearence than the current $J_z$, as expected. Movie~\ref{movie} reproduces the full temporal evolution of the current helicity field, highlighting the important establishment of the AR current structure at one given time (see below).

%
\begin{figure*}[h!] 
  \begin{center}
  \includegraphics[width=14cm]{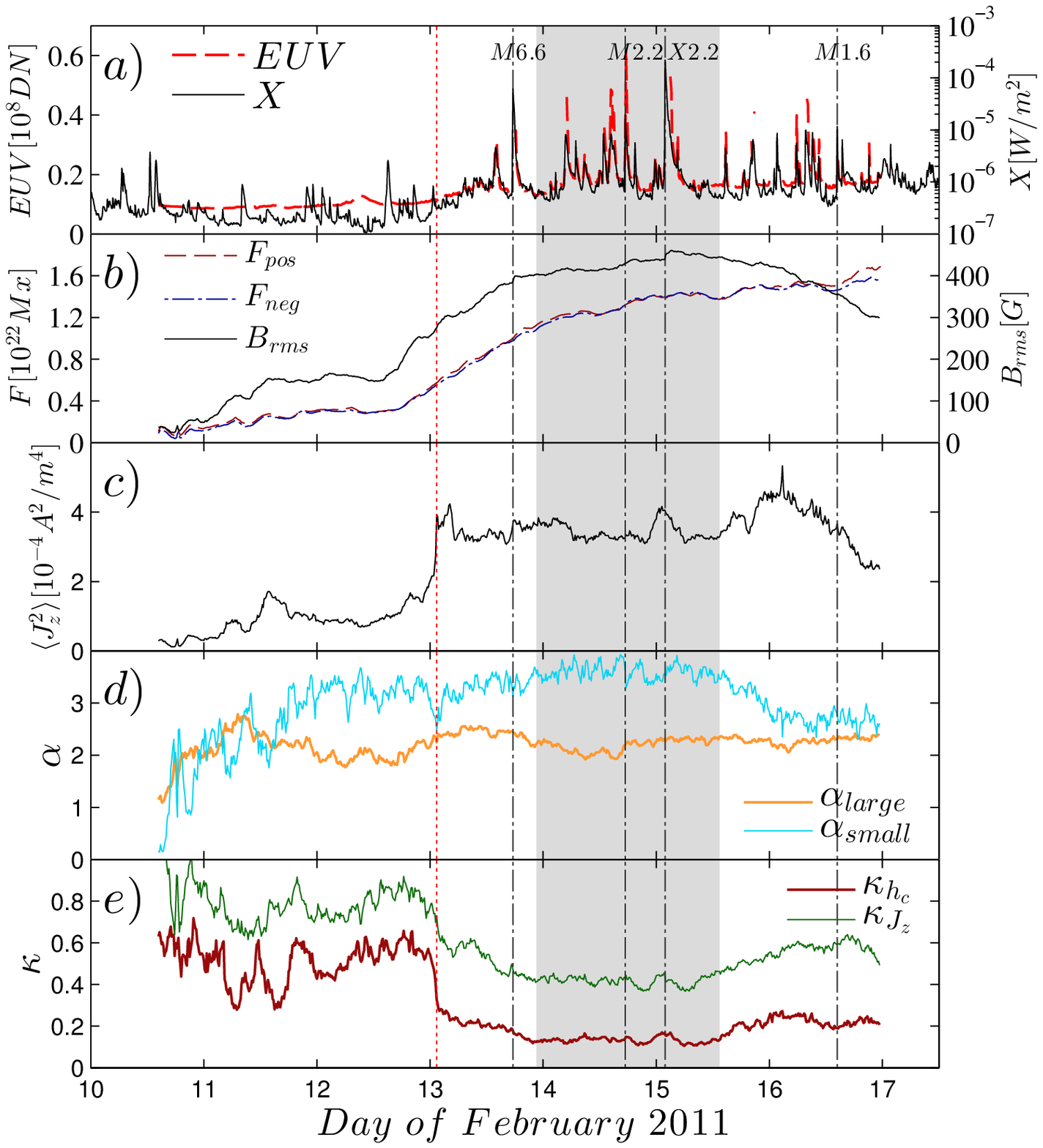}   
\caption{Time evolution of: (a) flux of X-ray (black full line) and EUV (red dot-dashed line); (b) space-integrated positive (thin, full blue line) and negative (thick, red dashed line) magnetic flux, together with the level of magnetic fluctuations (thick solid black line); (c) space-averaged squared vertical current $\langle J_z^2\rangle$; (d) spectral indices $\alpha_{small}$ (thin blue line) and $\alpha_{large}$ (thick yellow line); (e) cancellation exponents $\kappa_{J_z}$ (green thin line) and $\kappa_{h_c}$ (thick red line). Vertical grey dot-dashed lines indicate the time of the M and X flares. The red dashed vertical line indicates the transition time $t^\star$. The gray area is the stationary interval used for the statistical study.}
  \label{fig_temporal}
  \end{center}
  \end{figure*}

The amount of power connected to small scale magnetic fields can be quantitatively visualized through the typical fluctuation level of the magnetic field vector, $B_{rms}=\langle\delta B_x^2\rangle^{1/2} + \langle\delta B_y^2\rangle^{1/2} + \langle\delta B_z^2\rangle^{1/2}$, where $\delta B_i=B_i-\langle B_i\rangle$ are the fluctuations of the $i$-th magnetic field component, and the brakets indicate space average over the whole AR. Figure~\ref{fig_temporal}b shows the temporal evolution of $B_{rms}$, indicating the increase of fluctuations starting at the beginning of the emerging phase of the AR (see the magnetic flux), and then an evident decrease in the final part of the obsevation. 

Another useful quantity is the mean squared current $\langle J_z^2\rangle$, the average being again estimated over either the whole AR. This quantity is related to the level of dissipation of magnetic energy, and is customarily used in numerical simulations to describe the evolution of turbulence. Panel (c) of Figure~\ref{fig_temporal} shows the time profile of $\langle J_z^2\rangle$ (black solid line). From the figure, a sharp increase of $\langle J_z^2\rangle$ is seen at $t^\star=$1:24 UT on February 13th, during the initial stage of the flux emergence. This interestingly suggests that, unlike the smooth, slower and progressive emergence of magnetic flux and magnetic fluctuations, the transition toward a high mean vertical current state occurs in a very short time, $\Delta t^\star = 192$ minutes.

Finally, since we aim at connecting the magnetic properties of the AR with occurrence of flares, we have collected the measurements of the X-ray (X) and Extreme-Ultra-Violet (EUV) fluxes, as measured by GOES and AIA/SDO (131\AA~channel), respectively (Figure~\ref{fig_temporal}a).  

GOES provides the X-ray flux measurements integrated over the whole solar disc (black solid line). This could include in the measurement flaring activity from other active regions. However, we have checked that during the time interval under study (the gray shaded area in all panels of Figure~\ref{fig_temporal}), the X-ray flux was mainly originated in NOAA 11158. 

On the contrary, the EUV flux measurement retains the spatial information. Images taken at the 131\AA~band pass have been selected for their sensitivity to hot ($\sim10$MK) plasma to complement the GOES data. Level 1.5 data (cutout service) have been used at a 12-minute cadence. The maps have been integrated over the active region NOAA 11158, resulting in a time series of EUV emission from the area under study (red dot-dashed line in Figure~\ref{fig_temporal}). Some data gaps are present in the EUV time series. For the statistical analysis, such gaps have been filled through cubic spline interpolation of the data. During flaring time, the brightest parts of the flaring region are generally saturated. Therefore, the absolute value of the EUV emission does not correctly represents the intensity of the flare. Nevertheless, higher intensity might still represent larger flares, as generally large flares tend to have larger areas of saturated pixels. In any case, what is important for the study present in this paper is that the time series indicates flaring times properly. Therefore the saturated pixels are not of importance for the conclusions drawn in this paper. Moreover, the correspondence with flares eruptions estimated through the X-ray flux avoids any ambiguity.

Figure~\ref{fig_temporal}a shows that, after a quiet interval in the early stage of the AR prior to the flux emergence, both X and EUV flux background level sensibly increases, starting approximately at the transition time $t^\star$, in agrement with the increase of the dissipation proxy $\langle J^2\rangle$. 
After this time, several flares were released by NOAA 11158, as evident from the peaks in the temporal profile of both X-ray and EUV flux. These include 18 C-class, three M-class and one X2.2-class flares. Vertical dashed lines identify the exact time of eruption of the M and X flares in all panels of Figure~\ref{fig_temporal}.
%
%

%
%
\section{Scaling laws in NOAA 11158: spectral and cancellation analysis}
\label{sec-results}

\subsection{Spectral analysis}
 
\begin{figure}[h!] 
  \begin{center}
  \includegraphics[width=12cm]{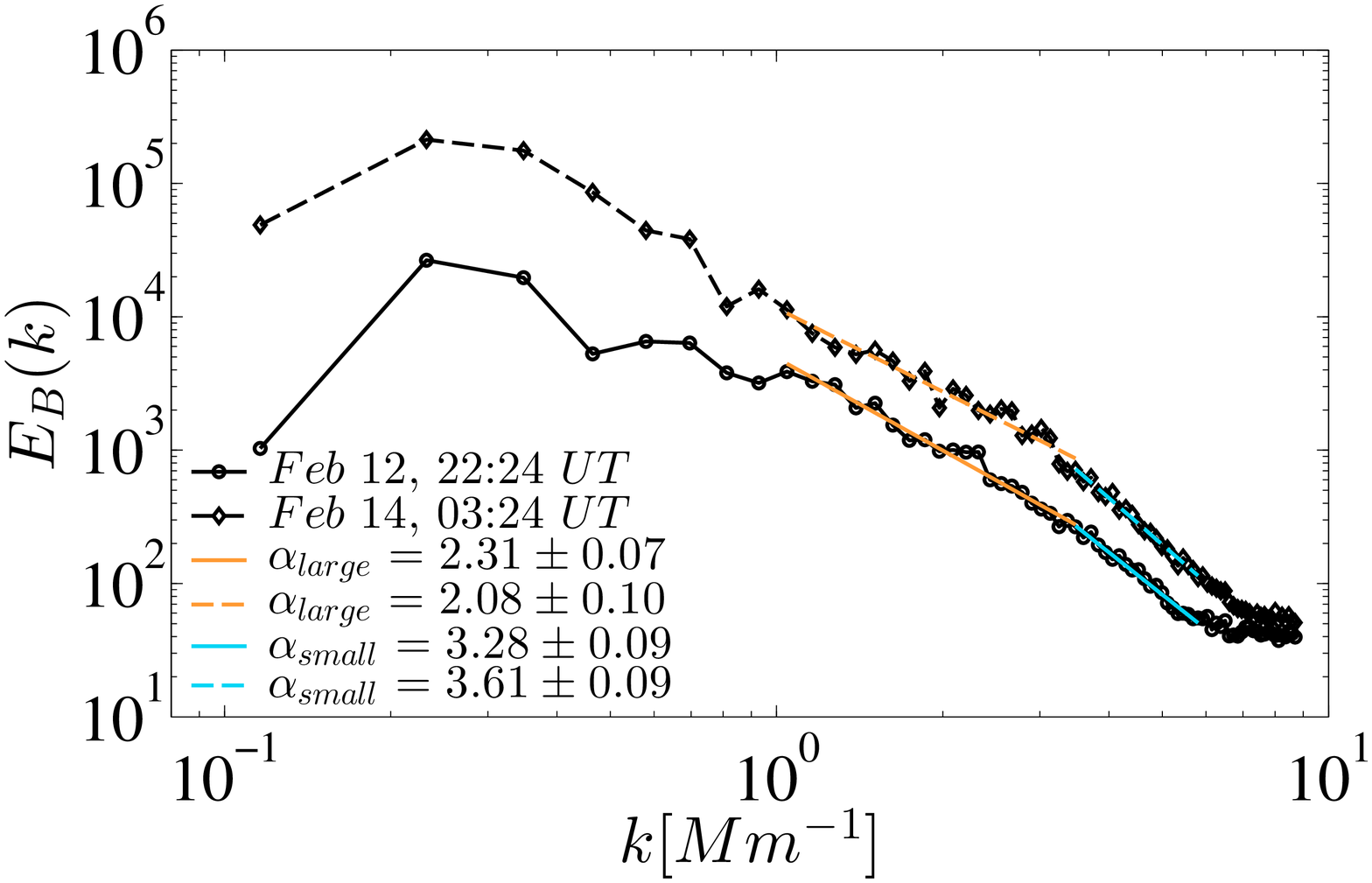}   
\caption{Magnetic power spectra estimated on February 12th at 22:24UT (circles) and on February 14th at 03:24UT (diamonds). Power law fits and the corresponding spectral indices are indicated.}
  \label{fig_spectrum}
  \end{center}
  \end{figure}

In order to characterize the possible presence of turbulent-like behaviours in the active region magnetic field, as suggested by the large mean current density, for each snapshot of the time series we have calculated the (omnidirectional) magnetic spectrum 
$E_B(k)=\int_{|\mathbf{k}|=k} |B(\mathbf{k})|^2 d\mathbf{k}$, where $k$ indicates the wavevector~\citep{zhang2014}. Examples of magnetic spectra are shown in Figure~\ref{fig_spectrum}, at the same two times as in Figure~\ref{fig_AR}, suggesting the presence of a power-law decay in the intermediate range of scales (the inertial range), approximately between $1.8$Mm and $6$Mm. 
%
%
%
In this range, a power-law fit $E_b(k)\propto k^{-\alpha_{large}}$ provides the scale spectral index $\alpha_{large}$. At the early stage of the AR emersion, the spectral power is smaller (Figure~\ref{fig_spectrum}), the power-law is less defined and the spectral index is very variable. At later times $t>t^\star$, when the active region is emerging, the scaling exponent is more stable, $\alpha\simeq 2$. This behaviour is clearly visible in the supplemental material~\ref{movie}, where the movie reproduces the time evolution of the spectra. The observation of a power-law spectrum, with spectral index compatible with a Kolmogorov-like phenomenology, confirms once again that the AR magnetic fields can be studied in the framework of turbulent flows, as already suggested in the past~\citep{abra-turbo}. 
At larger wavevectors, the spectrum is compatible with the presence of a secondary, different power-law, with $\alpha_{small} \simeq 3.3$, although the range of scales is rather limited. This behaviour indicates the presence of a characteristic scale, around $1.8$Mm, where the physical processes change. Note that recent estimaties of the Batchelor integral scales in the quiet solar photosphere provided similar values~\citep{abra-scales}. The presence of a spectral break is commonly observed in plasmas, for example at the transition from the magnetohydrodynamic range and the kinetic scales, where particle effects are not negligible and produces variations in the energy cascading mechanisms, resulting in steeper spectra. 
Examples of spectral break followed by steeper spectra are observed in solar wind plasmas~\citep{leamon,olga,fouad}, as well as in a variety of numerical simulations~\citep{hall,karimabadi}. 
At very large wave vectors the spectrum flattens, suggesting that scales $\lesssim 1$Mm may be affected by instrumental noise.
These spectral properties are roughly in agrement with previous analysis of the same active region~\citep{georgoulis2013,zhang2014,guerra2014} and of other ARs~\citep{abra-spectra} where, however, only inertial range spectal indices were evaluated.

\subsection{Cancellation analysis}

\begin{figure}[h!] 
  \begin{center}
  \includegraphics[width=13cm]{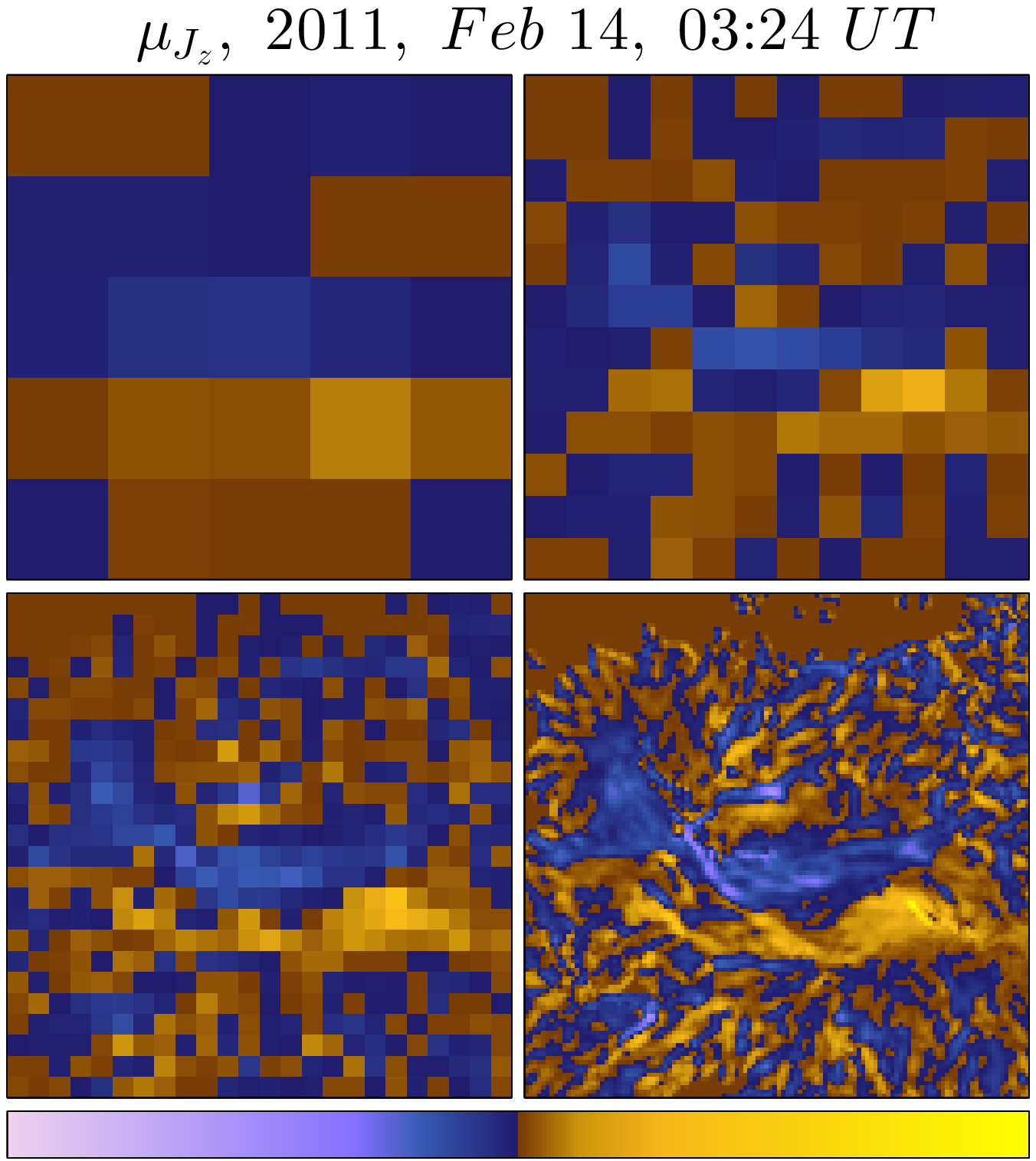}   
\caption{The signed measure $\mu_i(l)$ calculated for the vertical current density $J_z$ on February 14th at 03:24UT. Color scale is arbitrary.}
  \label{fig_mu}
  \end{center}
  \end{figure}

\begin{figure*}[h!] 
  \begin{center}
  \includegraphics[width=12cm]{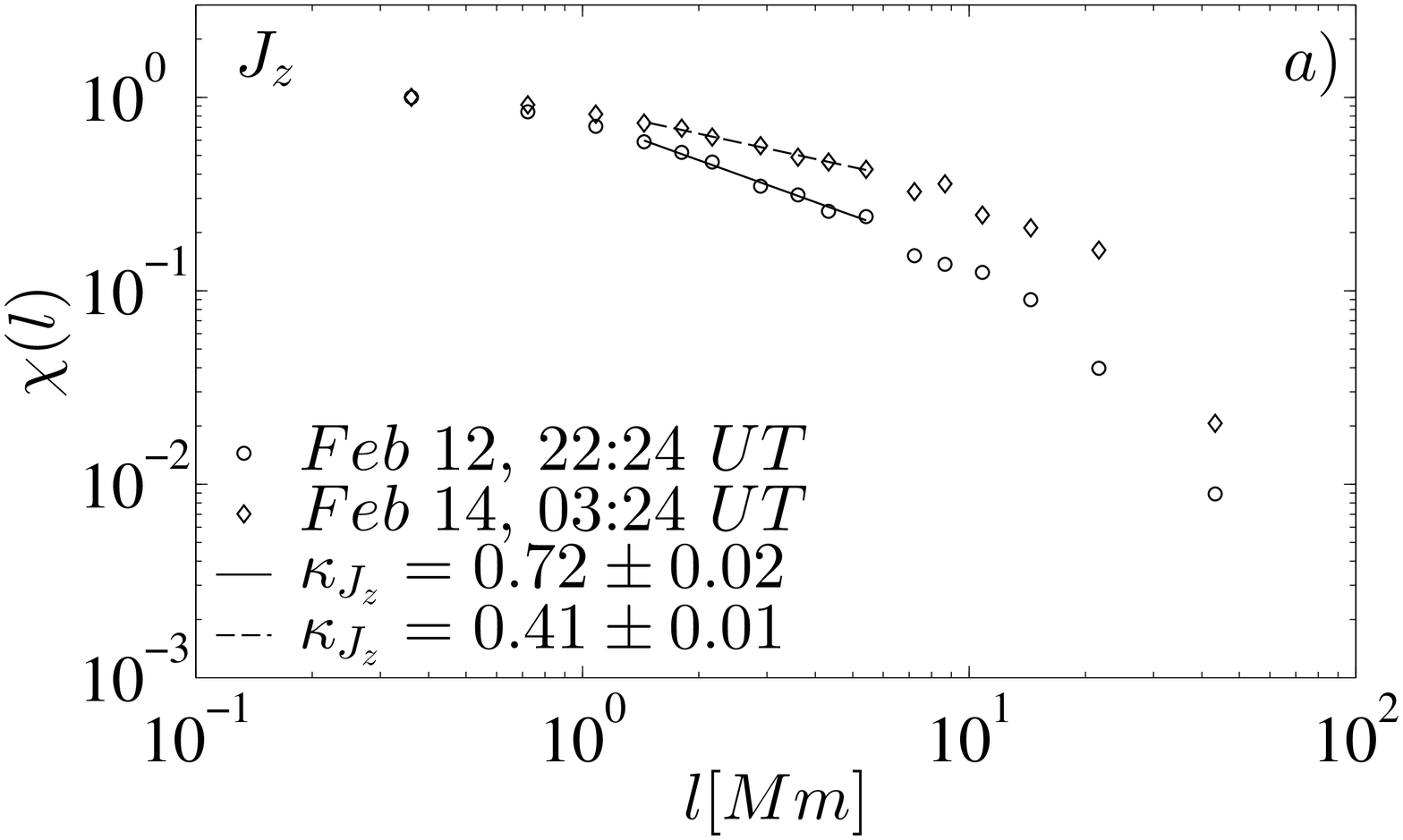} \\
  \vskip 12pt
  \includegraphics[width=12cm]{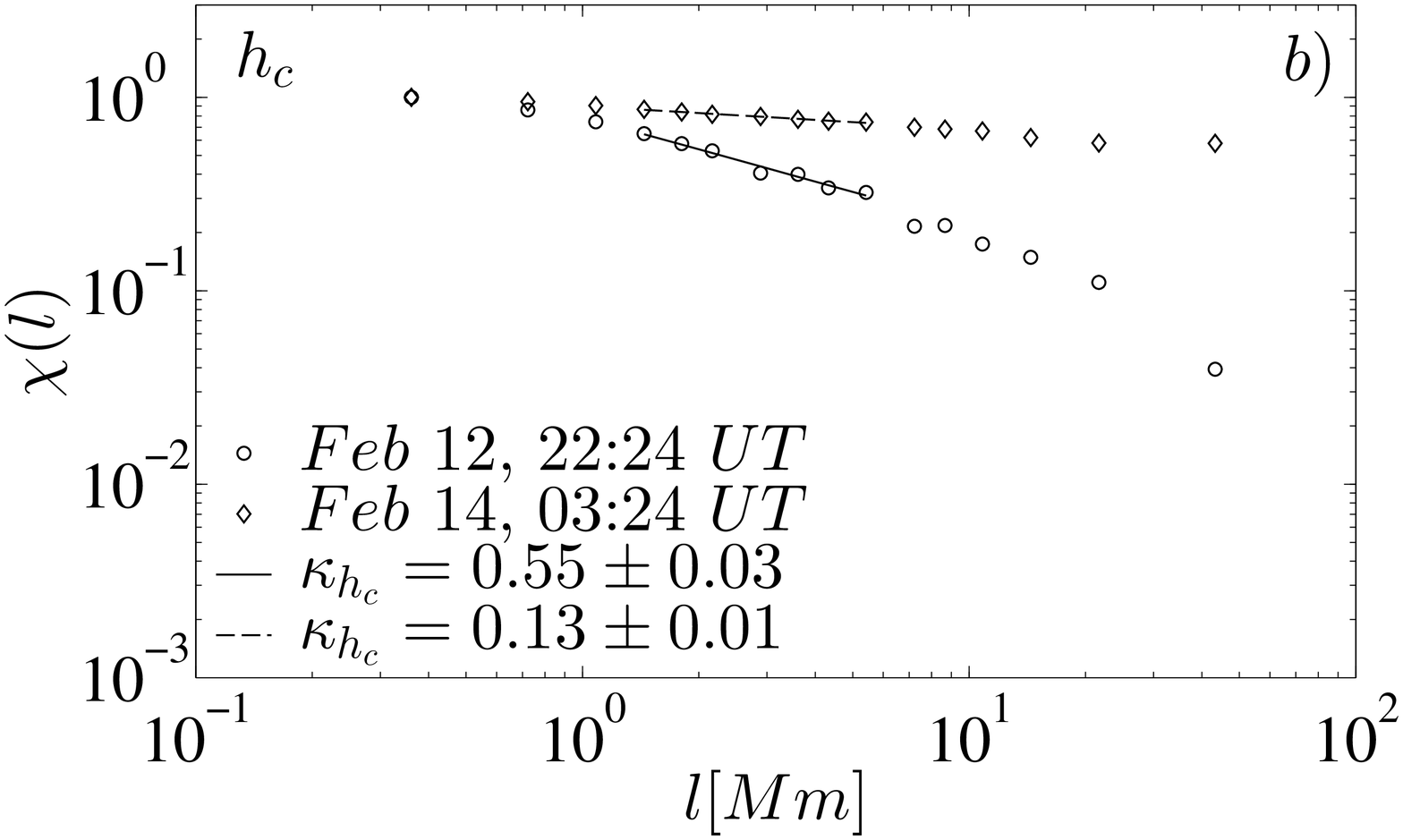}  
\caption{The scaling of the cancellation function $\chi(l)$ calculated (a) for the vertical current density $J_z$ and (b) for the current helicity $h_c$, on February 12th at 22:24UT (circles) and on February 14th at 03:24UT (diamonds). Power-law fits are also indicated, along with the corresponding cancellation exponents.}
  \label{fig_cancellation}
  \end{center}
  \end{figure*}

Upon confirmation that the active region magnetic field can be described in the framework of turbulence, cancellation analysis has been performed on the vertical current $J_z$ and on the reduced current helicity $h_c$, for each snapshot of the time series. Note that the magnetic field is relatively smooth, so that its cancellation analysis does not provide information about the dynamics of the AR~\citep{abramenko}.
Figure~\ref{fig_mu} shows one example of the signed measure maps, calculated from the data through equation (\ref{eq_mu}) at four different partition scales. 
%
%
While at large partition scales the positive and negative fluctuations cancel each other, resulting in small values of the signed measure, at smaller partition scales the sign-defined structures emerge as brighter regions of the maps. Note also the sign-defined large structures in the umbral part of the sunspots, where the magnetic field direction is well-defined. The way structures influence the overall signed measure at different scales is resumed by the cancellation function, depicted (for the same two snapshots as in previus figures) in Figure~\ref{fig_cancellation}, for both the current (top panels) and the current helicity (bottom panels). A clear power-law range is visible in the intermediate range of scales, conservatively between $1.4$Mm and $6$Mm, roughly corresponding to the ``inertial range'' of the spectrum shown in Figure~\ref{fig_spectrum}. A power-law fit has been performed in such range, as indicated in Figure~\ref{fig_cancellation}, and the corresponding cancellation exponents have been evaluated. For the examples given here, exponents are steeper before the transition time, $\kappa_{J_z}=0.72\pm0.02$ and $\kappa_{h_c}=0.55\pm 0.03$, while after the transition they become shallower, $\kappa_{J_z}=0.41\pm0.01$ and $\kappa_{h_c}=0.13\pm 0.01$. Based on the phenomenological argument given in Section~\ref{sec-cancellations}, these values correspond to the presence, in the emerging stage, of broken current filaments with fractal dimension $D_{J_z}=0.56\pm 0.04$, and current helicity filaments, $D_{h_c}=0.9\pm 0.06$. At later stage, current filaments dominate, with fractal dimension $D_{J_z}=1.18\pm 0.02$, while well resolved, almost smooth current helicity structures are observed, $D_{h_c}=1.74\pm 0.02$. The faint saturation to $\chi=1$ at small scales shows that small scale features are rather smooth, probably becuase of the instrumental noise, and in agrement with the spectral observations. 
The extremely high quality of HMI-SDO data thus allows a very accurate estimation of the cancellation effect. 
The complete temporal evolution of the cancellation function is represented in the movie~\ref{movie}, where it is possible to confirm the good quality of the fit during the whole AR observation, and the clear changes occurring during the AR evolution.
%
%
As a stricking difference with previous results, sign singularities are measured not only in the smoother current helicity, but also directly on the current density. This is the first observation of this kind, witnessing the excellent quality of the data, and providing further information on the fields dynamics. Note also that, while earlyer results on AR cancellation analysis were affected by limited space resolution~\citep{abramenko,yurchyshyn,prelude}, a more recent study of Hinode magnetic fields has shown a similar excellent spatial resolution~\citep{hinode}, but was still lacking full time coverage at high temporal resolution.


\section{Results: time evolution of the magnetic field complexity and relationship with large flares}
\label{sec-time}
Previous results on cancellation analysis have suggested that the topological properties of active regions magnetic fields experience changes in correspondence to eruption of major flares~\citep{abramenko,yurchyshyn,prelude,hinode}. The first observations indicated that such abrupt changes may anticipate the flares by a fraction of a hour. However, the limited time resolution of the data used so far has been a serious obstacle to the detailed description of the time evolution of the magnetic field complexity. HMI-SDO data used here provide for the first time high time resolution, and full time coverage during almost the whole lifetime of the active region. This dataset represents a unique opportunity to study the dynamical properties of the magnetic complexity, and their relationship with occurrence of flares.
Previous studies of temporal evolution of AR NOAA 11158 focused on the properties of magnetic flux, potential, non potential and free energy, number of loops, magnetic helicity, and misalignment angle~\citep{sun2012,aschwanden,vemareddy,tziotziou,liu2012}.

\subsection{General temporal dynamical properties: turbulent dissipation and flares}

Looking at the gross behaviour of the parameters can give information about the general magnetic and energetic properties of the active region. To this aim, in Figure~\ref{fig_temporal} we have collected the time-dependent parameters of the active region, as measured by the instruments or calculated for this work. We have already discussed in previous sections the general properties of the time evolution of: (a) the EUV and X-ray flux, useful to identify the time of eruption of flares calssification (the time of eruption of three M and one X flares is indicated by vertical dot-dashed black lines); (b) the positive ($F_{pos}$) and negative ($F_{neg}$) magnetic fluxes of the whole active region, indicating the slow and steady emergence of magnetic flux, and the similar behaviour of the magnetic fluctuations level ($B_{rms}$); (c) the space-averaged squared vertical current $\langle J_z^2\rangle$, which shows the abrupt onset of turbulence at time $t^\star$, and a successive steady state (indicated as a gray area in the plot) preceding a final smooth decrease. 
The other panels of Figure~\ref{fig_temporal} show: (d) the spectral indices $\alpha_{small}$ (thin blue line) and $\alpha_{large}$ (thick yellow line), obtained from the power-law fit of the spectra as described in Section~\ref{sec-results} for both ranges of scale; (e) the cancellation exponents estimated from the power-law fit of the cancellation functions, for both the vertical current ($\kappa_{J_z}$, green thin line) and the reduced current helicity ($\kappa_{h_c}$, thick red line). The complete temporal evolution of the cancellation exponents can be observed in more detail in the movie~\ref{movie}.

The time evolution of the spectral indices, shown in panel (d), is only weakly affected by the AR dynamics. Indeed, the spectral exponent reaches a steady state at early times (about February 12th for the small range spectral index, and mid February 11th for the large scale exponent), one-two days in advance with respect to the flux emergence and the sharp jump observed in $\langle J_z^2\rangle$. While there is no evidence of later evolution for $\alpha_{large}$ (for example at the full emersion of the AR, or in correspondence with the enhancement of the flaring activity)~\citep{georgoulis2013,guerra2014}, the small scale spectral index seems to start increasing at $t^\star$, and then reaches a broad steady state, starting from the beginning of day 14th up to about 14 hours in day 15th (gray area in Figure~\ref{fig_temporal}). The larger values $\alpha_{small}\simeq 3.5$ indicate a steeper scaling exponent, which could be attributed to a time interval of more efficient transport and dissipation of turbulent energy. This spectral modification, to our knowledge observed here for the first time, is in agreement with the increase of X-ray and EUV background flux, as well as with the enhanced flaring activity. In either cases, while the large scale properties of the AR magnetic fluctuations do not seem to be affected by the flaring activity, the small scales are sensitive (although just weakly) to the erupting phase of the AR. This observation confirms that flares are most likely connected to small scale dynamics.

Finally, the time evolution of the cancellation exponents is shown in panel (e). In the early stage of the AR emergence, the current and current helicity structures are not well defined yet, the AR being dominated by noise. In these conditions, the cancellation functions show variable sign-singularity, so that the cancellation exponents are highly fluctuating. A similar behaviour was recently observed in direct numerical simulations for the study of the transition to turbulence in kinetic dynamics of plasmas~\citep{devita,karimabadi}, and is due to the presence of not yet fully developed structures. At $t^\star$, an abrupt decrease of the exponents synchronizes well with the sharp jump of the averaged squared current density. This indicates, once again, that the complexity of the magnetic field changes in a very short time, according to the fast increase of currents in the AR. This change is followed by a slower, steady decrase, lasting about one day. Then, the two cancellation exponents reach a steady state, with values $\kappa_{J_z}\simeq 0.42$ and $\kappa_{h_c}\simeq 0.13$, typical of smoother structures expected in a highly dissipating plasma (gray area in the plot). After about 36 hours, the cancellation exponents start growing again, indicating an increase of complexity of the AR magnetic fields.

The whole gross temporal evolution of the AR can therefore be resumed as follows: (1) the eary stage of the AR, with no flaring activity, showing randomly emeging, disrupted current filaments associated with weak turbulent energy; (2) the setup of the flaring activity and magnetic flux emergence, marked by the sharp onset of turbulence occurring at $t^\star$, with stabilization of the structures, and the following day of steadier change in the paramenters; (3) a period of strong flaring activity, associated with enhanced level of magnetic fluctuations and turbulent dissipation, and charaterized by a steady state of the structures geometry (gray area in Figure~\ref{fig_temporal}); (4) finally, the weakening of the flaring activity, associated with the steady decrease of turbulence level, weakening of the dissipation (shallower small scale spectrum), and increase of the magnetic complexity. This final step of the AR temporal evolution seems to indicate the transition to a different state, where finer magnetic structures can build up without necessarily result in large flares, suggesting an improved capacity of energy storage at smaller scales. The study of this part of the evolution is however not going to be discussed any further in this paper.
This scenario is fully consistent with the association between eruption of flares, and the general properties of the AR dissipation and small scale magnetic complexity.

\subsection{Short time-scale features at flaring times}

Besides the gross dynamical behaviour of the AR, the high temporal resolution of HMI-SDO allows for the first time the detailed analysis of short time-scale features, which are extremely suitable for the study of rapid phenomena such as solar flares. The instrumental cadence of 12 minutes is in fact comparable with the typical time scales associated to the flaring process. 
In order to study the possible relationship between flares and magnetic turbulence properties, already pointed out in the recent years~\citep{prelude}, the study can be focused on the most intense flares occurred in NOAA 11158. Previous studies have indeed shown that mostly major flares can be associated with variations of the magnetic complexity, as can be expected. In Figure~\ref{fig_temporal}, four vertical lines indicate the time of eruption of the largest flares of NOAA 11158. As already mentioned, these are three M-class (one of which considerably larger than the others) and one X-class flare. 

\begin{figure*}[h!] 
  \begin{center}
  \includegraphics[width=15cm]{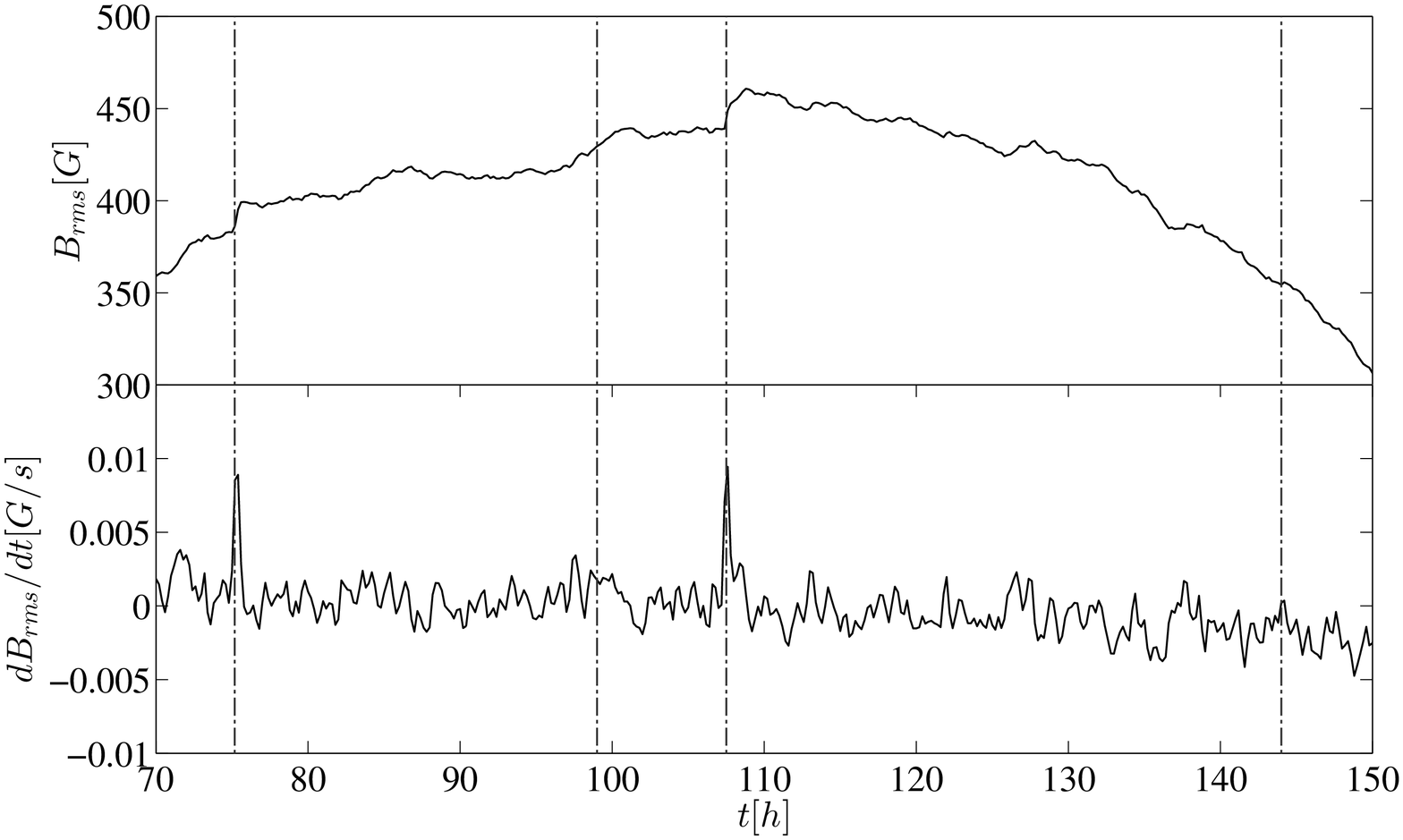} \\
  \caption{Top panel: the time evolution of $B_{rms}$ in the time interval including three M and one X class flares. Vertical bars represent the instant of major flares eruptions. Time is given in hours, starting at the beginning of the whole observation. Bottom panel: time derivative  $dB_{rms}/dt$, highlighting the sharp changes occurred at the major flares.}
  \label{fig_derivata}
  \end{center}
  \end{figure*}

It is evident from the panels (b) and (d) of Figure~\ref{fig_temporal} that, at the time of the flares, no specific features are observed in the magnetic flux and in the magnetic spectral exponents. However, finer observation of such and other similar parameters evaluated in the flare triggering regions have revealed small but significative changes~\citep{petrie2013,song2013}.   
On the contrary, the magnetic fluctuations $B_{rms}$ show a clear sudden increase at the times of the first M and X2.2 flares. This is highlighted in Figure~\ref{fig_derivata}, where a magnification around the M and X flaring periods is shown (panel a). In the same figure, the time derivative of the two fields are also shown, with the evident peaks at the first M and at the X flares. No features are visible for the second and fourth M flares~\citep{petrie2013,song2013}. 
At the same times, the mean squared current $\langle J_z^2\rangle$ also exhibits interesting features. The M-class flare is associated to a jump in the mean dissipation, while at the X-class flare an evident broad peak is present. In the latter case, the turbulent dissipation level increases by $27\%$ (two orders of magnitude larger than the $0.2\%$ relative standard deviation evaluated in the steady period preceding the flare) in about $216$ minutes, peaks about $\sim48$ minutes before the flare, and then decreases back to its steady value in about the same amount of time~\citep{song2013}. 
Similar features are also observed at the X flare for both cancellation exponents, where a $6\%$ and $28\%$ growth is present for the current density and current helicity indicators, respectively. These increases are larger, by one order of magnitude, than the $0.5\%$ and $0.8\%$ standard deviation levels in the steady period preceding the flare. The duration of the increase and decrease phases is approximately the same as for $\langle J_z^2\rangle$, i.e. of the order of 200 minutes. The same kind of behaviour seems to hold for the weaker features observed at the time of the first M-class flare for the current cancellation exponent, although these changes are not as evident as for the X2.2 flare, and are absent in the current helicity exponent. 
%
%

The increase of the mean squared current and of the magnetic complexity (as estimated through the cancellation exponents), which represent the main result of this work, can be interpreted as follows. During that phase, the AR is in a highly turbulent state (as shown by the spectra), with a steady, high level of dissipation (shown by the slightly steeper small scale spectra with respect to the non-flaring stage, and by the higher background of X-ray and EUV flux), and associated with the presence of relatively smooth current and current helicity structures. Shortly before the flare, magnetic gradients and complexity increase, suggesting the injection of an excess of magnetic energy, which is not fully dissipated, but rather stored through the build-up of field complexity, with resulting enhancement of the current filamentation in the AR. This finally results in the conditions for flaring. After the flare explosion, the conditions come back to the steady state, while the typical level of magnetic fluctuations starts to decrease.

It should be mentioned that the cancellation exponent features shown here are opposite with respect to provious cancellation analysis, where complexity was observed to decrease before a large flare~\citep{abramenko,yurchyshyn,prelude,hinode}. This difference could be due both to the lack of good time resolution, which could prevent the correct temporal description of the parameters, and to the poor spatial resolution of previous analysis.


\section{Evaluating correlations}

\begin{figure*}[h!] 
  \begin{center}
  \includegraphics[width=8cm]{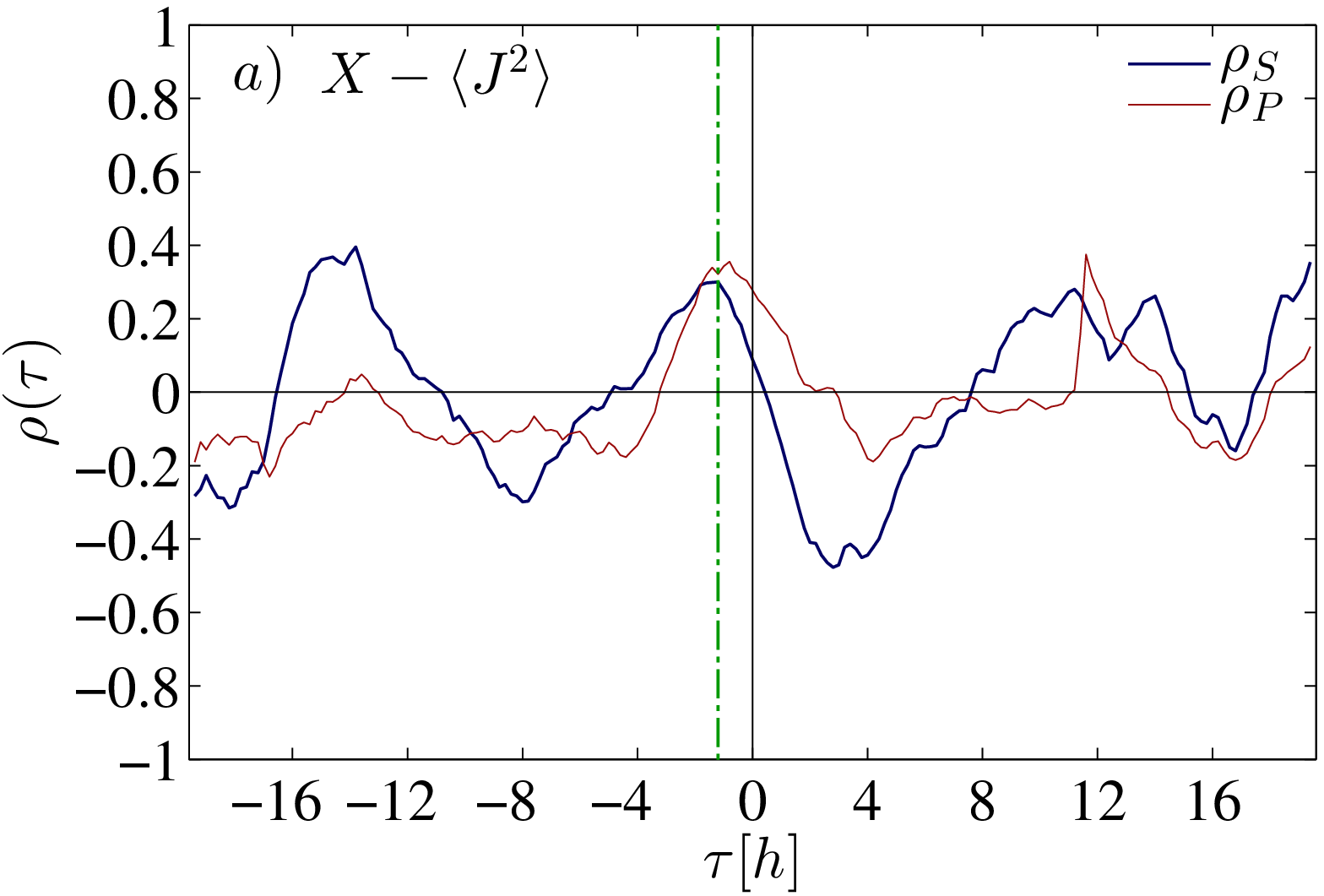} \includegraphics[width=8cm]{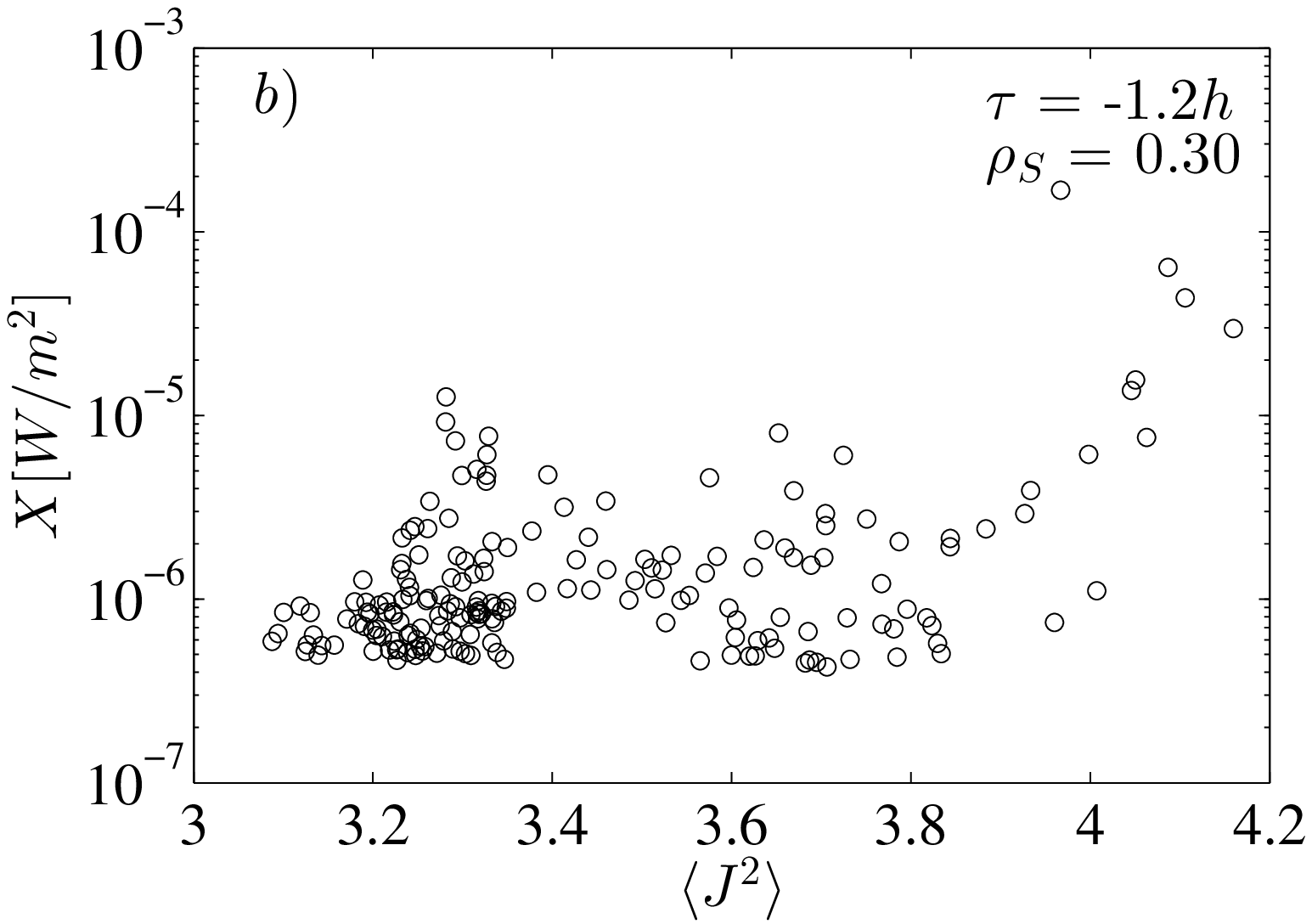}\\
  \includegraphics[width=8cm]{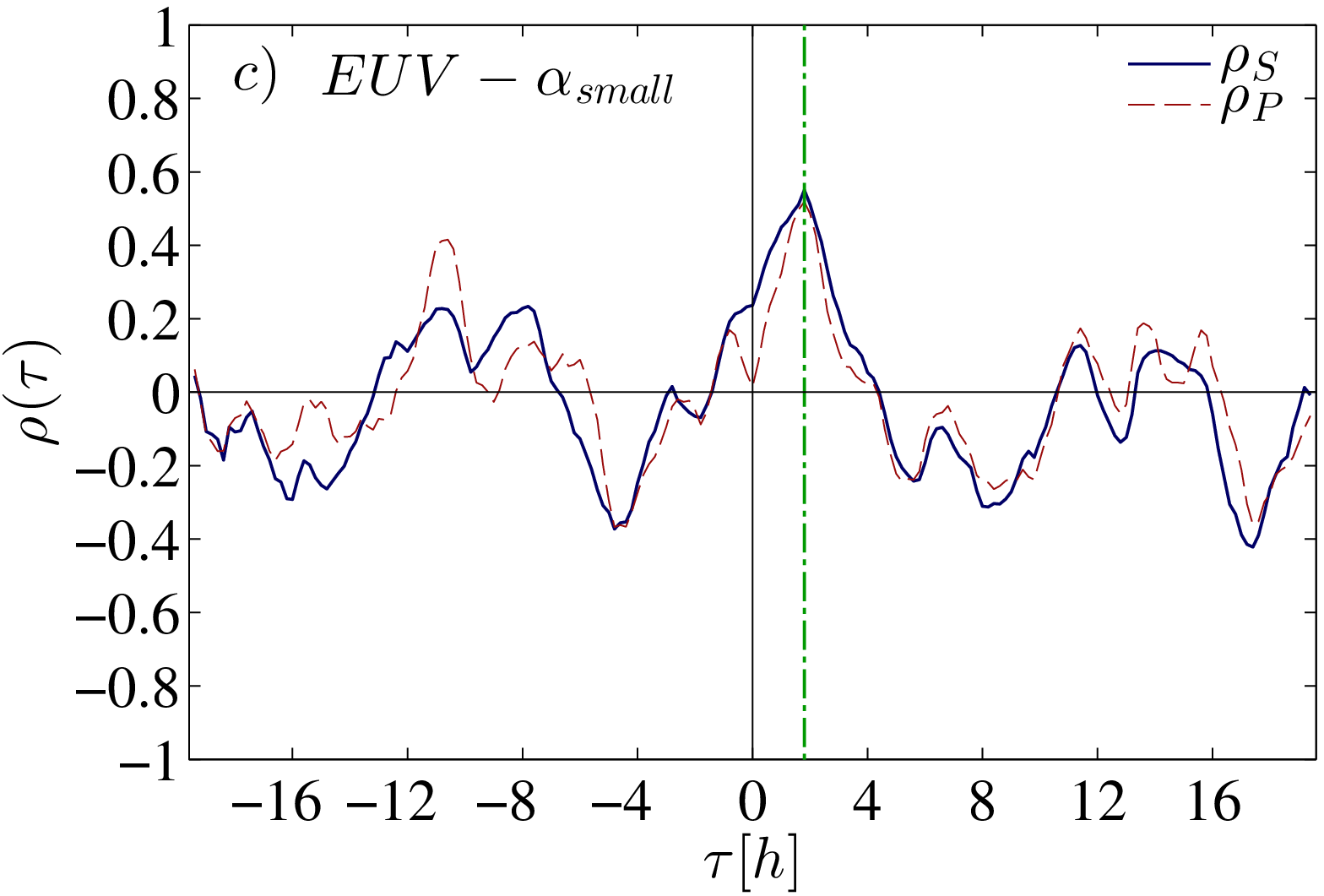} \includegraphics[width=7.5cm]{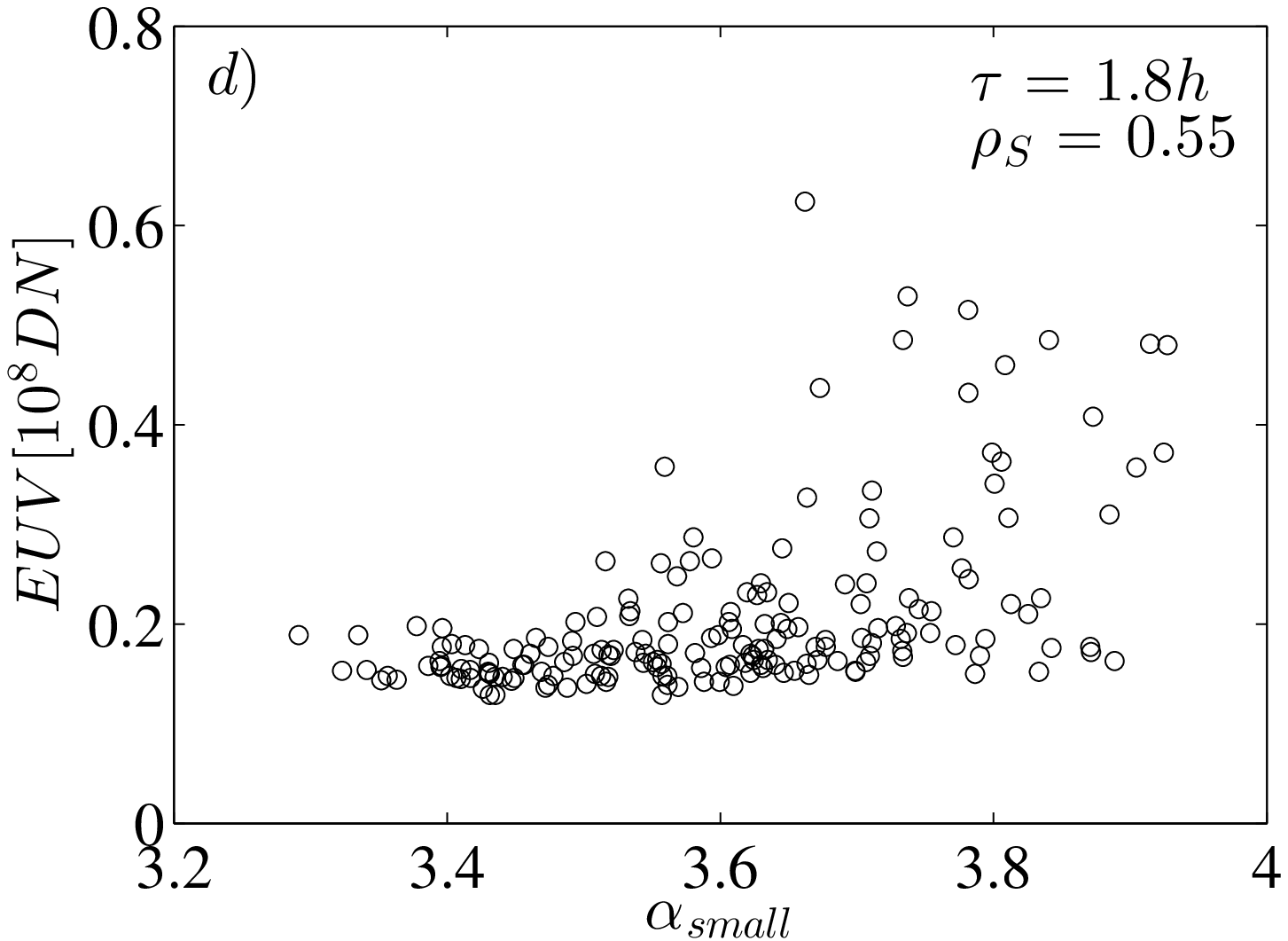}\\
  \includegraphics[width=8cm]{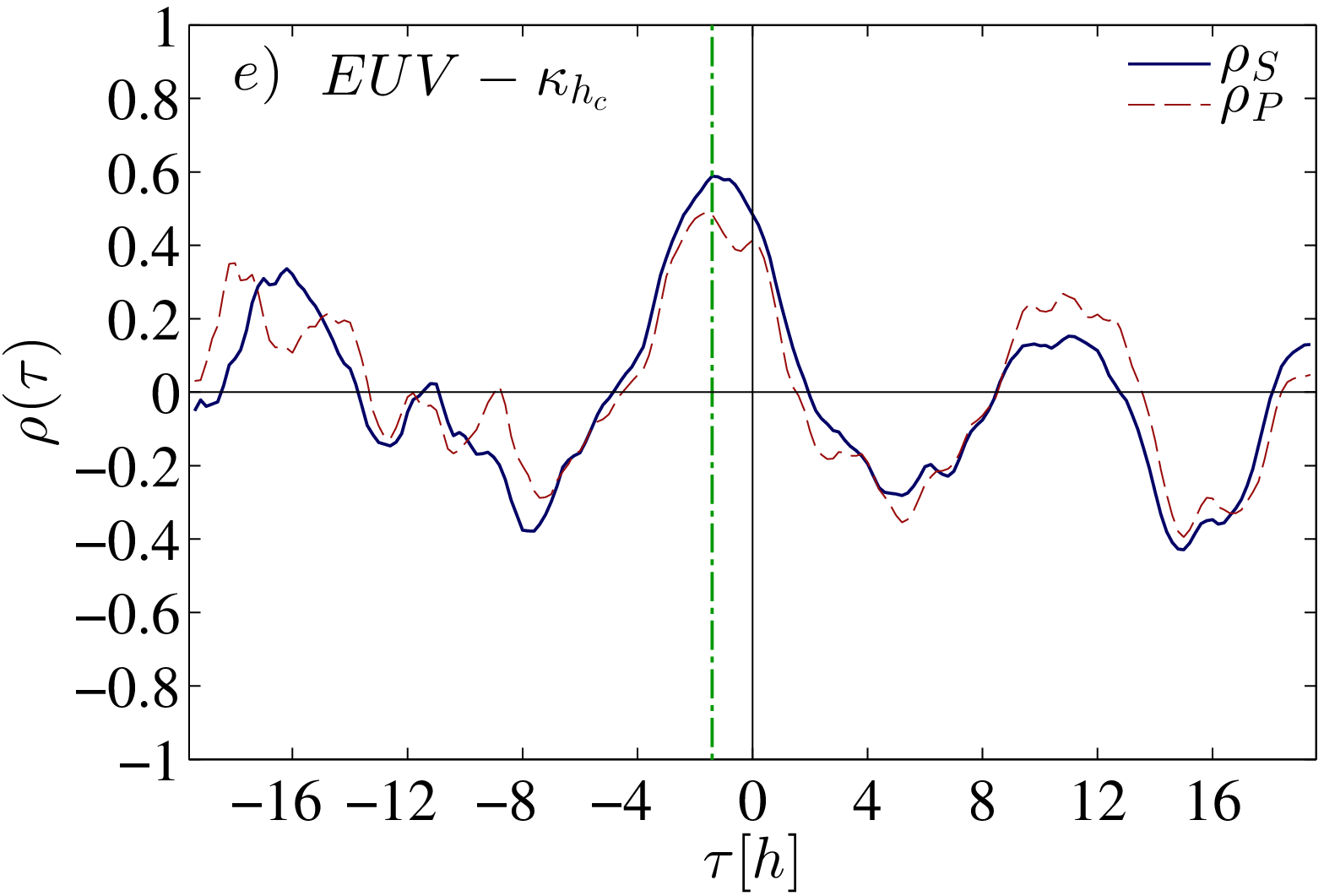} \includegraphics[width=7.5cm]{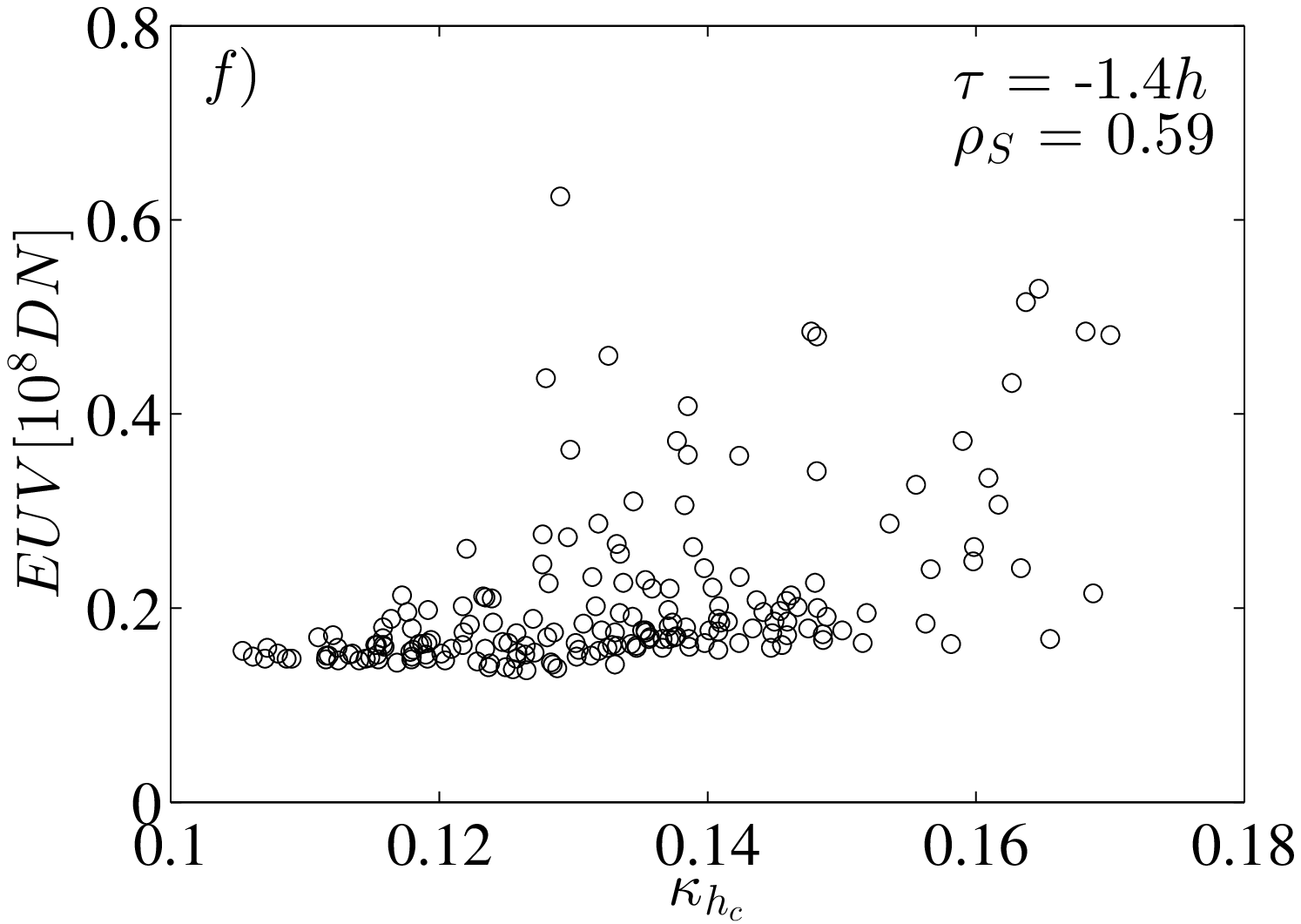}\\
\caption{Left panels: the correlation coefficients $\rho_P(\tau)$ and $\rho_S(\tau)$ at various time lags $\tau$, for the indicated paris of variables. The dot-dashed vertical line indicates the time of the maximum Spearman correlation. Right panels: the corresponding scatter plots at the time lag of the Spearman correlation peaks.}
  \label{fig_correlation}
  \end{center}
  \end{figure*}

The observation of the temporal behaviour of the active region magnetic structure has revealed the correspondence between changes in magnetic field properties, and the eruption of large flares. This is visible both in the gross evolution of the parameters and in the short time-scale features, as shown in Figure~\ref{fig_temporal}. 
In order to give a quantitative measure of the relationship between the observed features and the occurrence of flares, a statistical study of the correlations between the different parameters presented in Figure~\ref{fig_temporal} can be performed. This is possible thanks to the high time cadence of the data, which allows for the first time a significant statistical study. In order to limit the effect of transients, noise and smaller flares, we performed the correlation analysis only in the stationary stage of the activity, indicated by the gray area in Figure~\ref{fig_temporal}.
The cross-correlation coefficients between several pairs of parameters has been computed, and results are collected in Table~\ref{table_correlation}. We have used both the Pearson and the Spearman (ranks) cross-correlation coefficients, evaluated at the time lag where they are larger (in Table~\ref{table_correlation}, time lag of $0$ hours indicate absence of significative correlation). The Spearman coefficient is less sensitive to nonstationarity of the samples, and to nonlinearity of correlations. Since there is no particular reason to expect linear correlations, we believe that Spearman coefficient could be a more efficent parameter accounting for correlations in this system. 
Examples of correlation functions (for both ordinary and Spearman ranks correlations) are shown in left panels (a, c, e) of Figure~\ref{fig_correlation}. Peaks are evident for the current helicity cancellation exponents and for the small scale spectral index, while correlations are poor for the mean squared vertical current. The dashed vertical line indicates the time lag at which the Spearman correlation is maximum. 
As can be seen in Table~\ref{table_correlation}, while some of the pairs are not correlated, for others it is possible to highlight a relevant correlation. For example, EUV flux exhibit a strong degree of correlation $\rho_S=0.59$  with the current helicity complexity $\kappa_{h_c}$ (at $\tau=-1.4$ hours lag), or $\rho_S=0.55$ with the small scale spectral index $\alpha_{small}$ (at $\tau=1.8$ hours lag).  Note that maximum correlation time lags are negative, indicating that the changes in the parameters anticipate the X-ray and EUV emission, except for the spectral indices, which on the contrary react after the flares. The sign of the correlation suggests the causality direction of our observations. Increase of magnetic complexity and gradients anticipates the flares, while the dissipation becomes more efficient after the flares.
The most relevant correlation is found between EUV flux and current elicity cancellation exponent. This confirms that sign-singularity analysis is a suitable, sensitive tool, able to capture the fine variations in the AR magnetic complexity preceding the eruption of large flares.

Finally, more details about the nature of the observed correlations can be evidenced by showing the corresponding scatter plots for each pair of parameters, spaced with the time lag of the Spearman correlation peaks. These are shown in the right panels (b, d, f) of Figure~\ref{fig_correlation}. 
It appears evident that most of the correlation comes from the large flares (the top part of the scatter plots), while smaller X or EUV records (the background emission) are more randomly distributed with the other parameters.
Interestingly, all major flares occur above given values of the correlated variables, i.e. for the cancellation exponents and for the small scale spectral index. This suggests the presence of a threshold of the magnetic dissipation and topological complexity, below which no large flares are observed. This quantitative observation further confirms that flares are strictly related to small scale turbulent and dissipative processes in the photospheric magnetic fields, and that higher complexity of the currents enhances the probability of observing large flares.
%
%
\begin{table*}[ht]
\label{table_correlation}
\begin{center}
  \caption{Maximum correlation coefficients $\rho_P$ and $\rho_S$, and the time lags $\tau_P$ and $\tau_S$ (in hours), estimated between the X-ray and EUV fluxes and the indicated parameters.}
\begin{tabular}{ccccccccccc}
\hline
\multicolumn{1}{c}{ } & \multicolumn{2}{c}{$\kappa_{h_c}$} & \multicolumn{2}{c}{$\kappa_{J_z}$} & \multicolumn{2}{c}{$\langle J_z^2\rangle$} & \multicolumn{2}{c}{$\alpha_{large}$} & \multicolumn{2}{c}{$\alpha_{small}$}\\
   & EUV & X & EUV & X & EUV & X & EUV & X & EUV & X \\
\hline
$\tau_P[h]$  & -$1.60$ & -$0.80$ & -$1.40$ & -$1.40$ & -$1.60$ & -$1.40$ &  $1.00$ &  $1.20$ &  $1.80$ &  $2.40$ \\
$\rho_P$     &  $0.49$ &  $0.33$ &  $0.32$ &  $0.19$ &  $0.36$ &  $0.34$ & -$0.18$ &  $0.13$ &  $0.52$ &  $0.25$ \\
\hline
$\tau_S[h]$  & -$1.40$ & -$1.40$ & -$1.40$ & -$1.20$ & -$1.20$ & -$1.20$ &  $0.60$ & -$0.80$ &  $1.80$ &  $1.80$ \\
$\rho_S$     &  $0.59$ &  $0.50$ &  $0.31$ &  $0.19$ &  $0.27$ &  $0.30$ & -$0.27$ & -$0.19$ &  $0.55$ &  $0.49$ \\
\hline
\end{tabular}
\end{center}
\end{table*}
%
%

\section{Conclusions}

Motivated by the extremely good quality of HMI-SDO photospheric magnetic field vector measurements, we have studied some magnetic properties of the solar active region NOAA 11158, and their relationship with eruption of large flares (M and X class).
We have observed that the dynamics of magnetic fluctuations and mean vertical current describes quite well the transition of the AR into the flare activity stage. The spectral properties of the magnetic field fluctuations were also studied, suggesting the presence of a double range of scales. In the larger scales range, approximately corresponding to the inertial range of turbulence, the spectral properties are steady during most of the AR lifetime, and no particular features are observed at the time of major flares. On the contrary, the small scale range is characterized by a variable spectral index, which shows nontrivial correlations with the X-ray and EUV emission. In particular, the spectral slope increases (indicating more efficient dissipation of turbulent energy) during the central part of the observation, when flaring activity is enhanced. 
The magnetic field complexity was studied by means of the cancellation analysis of its sign-singularities, as evidenced through the study of the vertical current and of the reduced current helicity. Cancellation analysis applied to the AR provided the qualitative estimation of the fractal dimension of the current structures, which is reasonably steady during the flaring time interval. However, interesting peaks are observed in correspondence of the largest flare of the AR (X2.2), and to some extent also for the M class flares.
The results shown here suggests that it is possible to quantitatively measure the magnetic complexity evolution of the AR during flares, which support the scenario of increasing entanglement of current and magnetic field one-two hours ahead of big flares. 

{\bf Any comments on the consequences on the main flaring models?
Should we mention that there is no need to focus at the location of the flare, as necessary for other observables (see literature)?}.

[....]


 \begin{acknowledgements}
The research leading to these results has received funding from the European Community Seventh Framework 
Programme (FP7/2007-2013) under grant agreement n. 269297/``TURBOPLASMAS". 
 \end{acknowledgements}

\end{document}